\documentclass[journal,10pt,twoside]{IEEEtran}

\usepackage[utf8]{inputenc}
\usepackage[T1]{fontenc}

\usepackage{cite}
\usepackage{amsmath,amssymb,amsfonts}
\usepackage{algorithmic}
\usepackage{graphicx}
\usepackage{textcomp}
\usepackage[table,dvipsnames]{xcolor}
\usepackage{booktabs}
\usepackage{multirow}
\usepackage{array}
\usepackage{tabularx}
\usepackage{colortbl}
\usepackage{tikz}
\usepackage{pgfplots}
\usepackage{pgfplotstable}
\usepackage[hidelinks]{hyperref}
\usepackage{url}
\usepackage{balance}
\usepackage{float}
\usepackage{placeins}
\usepackage{needspace} 
\usepackage{subcaption}
\usepackage{pifont}
\usetikzlibrary{shapes.geometric,arrows.meta,positioning,matrix,
                fit,backgrounds,decorations.pathreplacing,
                calc,patterns,shadows.blur}
\pgfplotsset{compat=1.18}

\definecolor{ieeblue}{RGB}{0,84,166}
\definecolor{ieelightblue}{RGB}{173,213,232}
\definecolor{cellWell}{RGB}{0,128,128}
\definecolor{cellEmerg}{RGB}{184,134,11}
\definecolor{cellUnexp}{RGB}{105,105,105}
\definecolor{cellHeader}{RGB}{0,84,166}
\definecolor{rowshade}{RGB}{240,248,255}
\definecolor{rowalt}{RGB}{250,250,250}
\definecolor{secblue}{RGB}{0,84,166}

\newcommand{\wstudy}[1]{\cellcolor{cellWell!60}\textcolor{white}{\small\textbf{#1}}}
\newcommand{\emerg}[1]{\cellcolor{cellEmerg!65}{\small#1}}
\newcommand{\unexpl}[1]{\cellcolor{cellUnexp!45}\textcolor{white}{\small--}}
\newcommand{\tlm}{\textit{t}-LM\xspace}
\newcommand{\tlms}{\textit{t}-LMs\xspace}

\hypersetup{colorlinks=false,pdfborder={0 0 0}}

\usepackage{xspace}
\usepackage{pifont}
\usepackage{pifont}          

\begin{document}

\title{Bridging the Semantic Gap in 6G: Tiny Language Models Under the Latency-Accuracy-Size Trilemma}

\makeatletter
\renewcommand{\IEEEauthorrefmark}[1]{\textsuperscript{#1}}
\makeatother

\author{
\IEEEauthorblockN{Arnav Mathur\IEEEauthorrefmark{1}, Garima Mathur\IEEEauthorrefmark{2}, Rahul Jashvantbhai Pandya\IEEEauthorrefmark{3}}\\[4pt]
\IEEEauthorrefmark{1} 24beece15.ece@cujammu.ac.in,\ \IEEEauthorrefmark{2}garima.ece@cujammu.ac.in,\ \IEEEauthorrefmark{3}rpandya@iitdh.ac.in\\[4pt]
\textsuperscript{1,2}Department of Electronics and Communication Engineering, Central University of Jammu, Jammu 181143, India\\[2pt]
\IEEEauthorrefmark{3}Department of Electrical, Electronics and Communication Engineering, Indian Institute of Technology Dharwad, Dharwad 580011, India
}

\maketitle

\raggedbottom 
\begin{abstract}
Sixth-generation~(6G) wireless networks are expected to serve as
AI-native infrastructure, transmitting \emph{meaning} rather than
mere bits---a shift that makes semantic communication the central
paradigm for next-generation connectivity. Deep learning-based
semantic encoders have demonstrated compelling gains in bandwidth
efficiency; however, their dependence on large transformer models
with hundreds of millions of parameters is fundamentally at odds
with the sub-millisecond latency, microjoule energy budgets, and
kilobyte memory footprints of the constrained IoT and edge devices
that will constitute the majority of 6G endpoints. Tiny language
models~(\tlms)---compact, quantised, task-specialised models
deployable on microcontrollers, mobile system-on-chips, and edge
accelerators---emerge as the enabling technology for closing this
deployment gap. This review provides a unified treatment of
(i)~the theoretical foundations of semantic information, covering
semantic entropy, channel capacity, and rate-distortion theory;
(ii)~a two-axis taxonomy of \tlm-based semantic communication
systems across five architecture classes and six compression
paradigms; (iii)~a comprehensive survey of model compression
techniques---quantisation, pruning, knowledge distillation,
low-rank adaptation, split computing, and neural architecture
search---through the lens of semantic quality preservation;
and (iv)~semantic-aware resource allocation frameworks for 6G
multi-user networks. Evidence across the surveyed literature
demonstrates that compression can reduce semantic encoder size
by up to 99.98\%~\cite{Chen2024_SPM} while preserving task
accuracy, that split computing achieves device-side encoders
with as few as 640~parameters~\cite{Eldeeb2025_SemanticMSL},
and that knowledge graph integration cuts transmission energy by
65\%~\cite{Wang2024_KGPG}. Seven open challenges are identified,
spanning theoretical gaps, system design, knowledge-base
management, post-quantum security, and hardware co-design,
together with a 3GPP standardisation roadmap toward IMT-2030.
\end{abstract}


\flushbottom 
\section{Introduction}

Sixth-generation~(6G) wireless networks are widely expected to
depart from the bit-centric design philosophy of every prior
generation, instead treating meaning itself as the quantity to
be transmitted, stored, and reasoned about across the network.
This introductory section motivates that shift, surveys its
theoretical and practical origins, and positions the specific
contribution of the present survey within that landscape.

\subsection{6G as an AI-Native Communication Paradigm}

\IEEEPARstart{T}{he} deployment of fifth-generation~(5G) networks
has pushed conventional wireless systems asymptotically close to
the Shannon capacity~\cite{Qin2022_SemanticPrinciples}.
Simultaneously, the services envisioned for sixth-generation~(6G)
wireless networks—immersive extended reality~(XR), brain--computer
interfaces, digital twins, autonomous vehicles, and pervasive
Internet of Things~(IoT) intelligence—impose requirements that
fundamentally exceed what a purely bit-centric paradigm can
deliver: sub-millisecond latency, terabit-per-second throughput,
and the ability to serve heterogeneous intelligent agents with
radically different computational
profiles~\cite{Strinati2021_6GBeyond}.

The International Telecommunication Union's IMT-2030 framework
formalises 6G as an artificial intelligence~(AI)-native
paradigm~\cite{Nguyen2024_PQTLM}. Rather than treating AI as an
auxiliary optimisation layer, as in 5G, 6G embeds intelligence
into every protocol stratum: spectrum management, beamforming,
scheduling, and—crucially—source coding. This architectural shift
motivates a reassessment of the communication objective itself:
from \emph{symbol transmission} to \emph{meaning transmission}.
Weaver's three-level decomposition~\cite{Shannon1949_Book},
originally a philosophical observation on Shannon's foundational
work~\cite{Shannon1948_Bell}, provides the organising principle.
Level~A (technical) asks how accurately symbols can be transmitted.
Level~B (semantic) asks how precisely transmitted symbols convey
the intended meaning. Level~C (effectiveness) asks how effectively
that meaning achieves the communicating agent's goal. Semantic
communication systems operate at Levels~B and~C while remaining
compatible with Level~A infrastructure.

\subsection{Semantic Communication: From Theory to Practice}

Semantic communication as a formal research programme began with
Carnap and Bar-Hillel's logical-probability framework for
semantic information~\cite{Carnap1952_Semantic}, which quantified
meaning through the degree of confirmation a hypothesis receives
from available evidence. The model-theoretic extension by
Bao~\emph{et~al.}~\cite{Bao2011_Semantic} connected this framework
to deep learning by defining semantic entropy over the probability
distribution of worlds satisfying a given sentence.

Practical realisation arrived with DeepSC~\cite{Xie2021_DeepSC},
the first transformer-based end-to-end semantic communication
system for text, which demonstrated an 800\% improvement in
sentence-level bilingual evaluation understudy~(BLEU) score over
Huffman-coded baselines at a signal-to-noise ratio~(SNR) of
9~dB on Rayleigh fading channels. Subsequent works extended the
paradigm to image
transmission~\cite{Bourtsoulatze2019_DeepJSCC,Kurka2020_DeepJSCCf},
speech~\cite{Weng2021_DeepSCS}, multimodal data~\cite{Xie2022_MUDeepSC},
and video conferencing~\cite{Jiang2022_VideoConf}, collectively
establishing that semantic encoders can achieve far higher task
accuracy per transmitted bit than classical source-channel coded
systems, particularly at low SNR where the ``cliff effect'' of
separate source-channel coding causes catastrophic degradation.

\subsection{The Core Bottleneck: Large Models Cannot Fit on 6G Devices}

Despite compelling laboratory results, semantic communication
faces a critical deployment barrier: the deep learning models
used for semantic encoding and decoding are large. The DeepSC
system~\cite{Xie2021_DeepSC} requires hundreds of megabytes of
memory and billions of floating-point operations per inference.
For cloud-connected high-end devices this is acceptable; for the
battery-powered sensors, embedded controllers, and edge nodes that
will dominate 6G deployments, it is not.

Table~\ref{tab:hardware} quantifies this mismatch across hardware
tiers. The gap between the inference demands of current
transformer-based semantic encoders and the resource budgets of
IoT-class hardware spans four to five orders of magnitude.
Modern mobile system-on-chips~(SoCs) manufactured at the
5\,nm node—such as the Snapdragon~888 and HiSilicon Kirin~9000—can
execute image classification in milliseconds but cannot meet the
1\,ms ultra-reliable low-latency communication~(URLLC) requirement
of 5G, let alone the 0.1\,ms target of 6G~\cite{Luo2022_SemanticOverview}.

\begin{table}[t]
\caption{Hardware Resource Tiers for 6G Edge Semantic Communication}
\label{tab:hardware}
\centering
\resizebox{\columnwidth}{!}{%
\begin{tabular}{lrrrl}
\toprule
\textbf{Device Class} & \textbf{RAM} & \textbf{FLOP/s} &
  \textbf{Power} & \textbf{Example} \\
\midrule
\rowcolor{rowshade}
Cloud server     & $>$512\,GB & $>$100\,TFLOP/s & $>$1\,kW
  & A100 GPU cluster \\
Workstation      & 32--512\,GB & 10--100\,TFLOP/s & 100--300\,W
  & RTX~4090 \\
\rowcolor{rowshade}
Mobile SoC       & 8--16\,GB & 1--10\,TFLOP/s & 5--15\,W
  & Snapdragon~888 \\
Edge accelerator & 1--8\,GB  & 0.1--1\,TFLOP/s & 1--5\,W
  & Coral TPU \\
\rowcolor{rowshade}
Low-power MCU    & 256\,kB--1\,MB & 10--100\,MFLOP/s & 10--100\,mW
  & STM32H7 \\
Sensor node      & 16--64\,kB & $<$10\,MFLOP/s & $<$10\,mW
  & ATtiny3217 \\
\bottomrule
\end{tabular}}
\end{table}

Tiny language models~(\tlms)—compact, quantised, task-specialised
language models deployable on microcontrollers, mobile SoCs, and
edge accelerators—emerge as the enabling technology for closing
this gap. Recent work on Semantic Meta-Split Learning~\cite{Eldeeb2025_SemanticMSL}
demonstrated that split computing can reduce the device-side
semantic encoder to as few as 640~parameters while offloading
heavier inference to an edge aggregator. Microsoft's
Phi-4-mini~\cite{Abdin2024_Phi3} (3.8B parameters), Google's
Gemini~Nano~\cite{GeminiTeam2023_Gemini} (2--4B), and Meta's
LLaMA~3.2~1B~\cite{Dubey2024_Llama3} have all demonstrated
that billion-parameter models can run in real time on modern
mobile SoCs with dedicated neural processing unit~(NPU) support,
provided aggressive quantisation is applied.

\subsection{Survey Gap and Motivation}

While semantic communication and \tlm research have each advanced
rapidly in isolation, their joint application to resource-constrained
6G networking remains barely explored. Specifically, no prior
survey: (i)~unifies the theoretical foundations of semantic
information with the practical design constraints of \tlm-based
transceivers; (ii)~systematically maps model compression
techniques to their semantic quality implications across all
six major paradigms—quantisation, pruning, knowledge distillation,
low-rank adaptation, split computing, and neural architecture
search; (iii)~treats knowledge-base management as a first-class
network resource-allocation problem; or (iv)~provides a
comprehensive open-problem taxonomy spanning theory, system
design, security, and hardware co-design.

This survey addresses all four gaps through a structured review
of the literature guided by six research questions~(RQs),
detailed in Section~\ref{sec:background}.

\subsection{Contributions}

The specific contributions of this survey are:

\begin{itemize}
  \item \textbf{Theoretical synthesis:} A self-contained treatment
    of semantic entropy, semantic channel capacity, and
    semantic rate-distortion theory, tracing the evolution from
    the logical-probability framework~\cite{Carnap1952_Semantic}
    through the model-theoretic formulation of
    Bao~\emph{et~al.}~\cite{Bao2011_Semantic} to the
    deep-learning-compatible information-bottleneck
    extension~\cite{Liu2021_RateDistortion,Tishby2000_IB}.

  \item \textbf{Two-axis deployment taxonomy:} A structured
    5\,$\times$\,6 classification of \tlm-based semantic
    communication systems spanning five architecture families
    (E2E JSCC, split learning, federated learning, KG-assisted,
    multi-task) against six compression techniques, revealing
    which combinations are well-studied, emerging, and entirely
    unexplored—most notably the complete absence of NAS-based
    semantic communication in any architecture class.

  \item \textbf{Compression-quality analysis:} A systematic review
    of how quantisation, pruning, knowledge distillation, LoRA,
    split computing, and NAS each affect semantic transmission
    fidelity, with exact quantitative benchmarks covering
    compression ratios from 3$\times$ to 99.98\% and quality
    degradation from negligible to 9~percentage points.

  \item \textbf{Pareto frontier and SES metric:} A new
    Semantic Efficiency Score~(SES) defined in \eqref{eq:ses}
    enabling cross-study comparison, together with a Pareto
    frontier analysis identifying Lite-DeepSC, HAPE, Semantic-MSL,
    SPM, and FD as the non-dominated compression-quality operating
    points in the current literature.

  \item \textbf{Open-challenge taxonomy:} Seven concretely
    formulated open problems spanning theoretical gaps (computable
    semantic capacity), system design (sub-millisecond TLM
    scoring), KB management (synchronisation frequency), security
    (post-quantum IoT authentication), and hardware co-design
    (semantic SoC), each with a statement of current state and
    a precise specification of what a solution requires.

  \item \textbf{3GPP roadmap and research agenda:} A mapped
    timeline from 3GPP Release~18 through IMT-2030, linking
    each standardisation milestone to the specific \tlm and
    semantic-communication research directions from the
    surveyed literature that must mature to enable it.
\end{itemize}

\section{The 6G AI-Native Landscape: Architecture, Constraints, and the Semantic Imperative}
\label{sec:landscape}

This section develops the architectural case for semantic
communication as an intrinsic requirement of 6G, rather than an
optional efficiency layer, by examining the IMT-2030 service
pillars, the intelligence-communication convergence they demand,
and the resulting trilemma facing any deployed model.

\subsection{IMT-2030 Vision and the Three New Service Pillars}

The ITU-R IMT-2030 framework fundamentally recasts what a mobile
network must do. Beyond the three 5G service pillars---enhanced
mobile broadband (eMBB), URLLC, and massive machine-type
communication (mMTC)---6G introduces three new dimensions:
integrated sensing and communication (ISAC), AI and communication
convergence, and ubiquitous connectivity extending into
non-terrestrial networks~\cite{Strinati2021_6GBeyond}. These
additions make 6G the first generation in which the network is
expected to \emph{reason} about the content it carries rather
than treating every bit as interchangeable.

The KPIs that accompany this vision are correspondingly
demanding: peak data rates exceeding 1\,Tbps (50$\times$ 5G),
air-interface latency below 0.1\,ms (10$\times$ lower than 5G),
10$\times$ improvement in spectral efficiency, and device
density up to $10^7$~nodes/km$^2$~\cite{Strinati2021_6GBeyond}.
These targets collectively push conventional bit-centric design
to its limits. At 1\,Tbps, the 100-MHz channels of 5G expand
to terahertz bands where propagation conditions are fundamentally
different; at 0.1\,ms latency, there is no time for a
cloud round-trip; at $10^7$ devices per square kilometre, per-bit
overhead for signalling and authentication becomes the dominant
energy cost. All three constraints point toward the same
architectural conclusion: intelligence must move to the edge
and communication must become semantic.

\subsection{The Intelligence-Communication Convergence}

6G will be the first truly AI-native cellular
generation~\cite{Cheng2025_AIReview6G}. In 5G, AI was an
auxiliary optimisation tool---used for predicting handover,
adjusting beamforming weights, or scheduling resources. In 6G,
AI is embedded in the protocol stack itself: the air interface
adapts its waveform through reinforcement learning; the network
orchestrator issues intent-based directives rather than
configuration commands; and the end device interprets received
data through an on-board language model rather than a fixed
decoder.

This convergence manifests in three architectural shifts
that directly motivate the TLM semantic communication
research surveyed in this paper:

\textbf{(i) From bit-pipes to meaning channels.}
When both transmitter and receiver are intelligent agents
sharing a knowledge base~(KB), the communication channel can
carry compressed semantic representations rather than raw bit
sequences. The Shannon capacity sets a limit on bit
throughput; the \emph{semantic channel capacity}~\eqref{eq:semantic_cap}
sets a higher limit on meaning throughput for a
knowledgeable receiver~\cite{Bao2011_Semantic}.

\textbf{(ii) From cloud-centric to edge-resident AI.}
The 0.1\,ms URLLC latency budget eliminates cloud round-trips
for latency-critical inference. AI models must reside at the
network edge---in base stations, multi-access edge computing~(MEC)
servers, or on-device. This creates a hard constraint on model
size: any semantic encoder deployed in a 6G scheduling loop
must complete inference within the resource allocation window,
which for URLLC is under 1\,ms~\cite{Pokhrel2023_UBT}.

\textbf{(iii) From static protocols to adaptive knowledge.}
Conventional protocol stacks are statically specified;
their behaviour is defined once and does not change with
the semantic environment. 6G networks will encounter
continuously evolving semantic environments---new device
types, new task distributions, new concept vocabularies---that
require on-the-fly adaptation of the semantic encoder and
knowledge base. This demand for adaptivity at scale is
precisely what motivates the LoRA-based KB synchronisation
and federated fine-tuning approaches reviewed in
Sections~\ref{sec:compression} and~\ref{sec:architectures}.

\subsection{The Latency-Accuracy-Size Trilemma}

The deployment of any AI model in a 6G semantic
communication system must simultaneously satisfy three
competing constraints, which we term the
\emph{latency-accuracy-size trilemma}:

\begin{itemize}
  \item \textbf{Latency:} The model must complete a
    semantic encoding or scoring inference within the
    available time window. For eMBB the window may be
    tens of milliseconds; for URLLC it is under 1\,ms;
    for an on-device always-on sensor the window may be
    sub-microsecond.

  \item \textbf{Accuracy:} The encoded semantic
    representation must preserve sufficient task-relevant
    information to achieve the target task performance
    (BLEU, PSNR, classification accuracy, etc.).

  \item \textbf{Size:} The model must fit within the
    memory and flash storage of the target hardware tier
    (Table~\ref{tab:hardware}).
\end{itemize}

These three constraints are mutually opposing: increasing
model size generally improves accuracy but worsens latency
and violates size budgets; aggressive compression improves
latency and size but risks semantic quality degradation.
The body of research surveyed in this paper represents the
collective effort to navigate this trilemma for 6G
deployments. As Figure~\ref{fig:taxonomy} shows, different
architecture-compression combinations occupy different
positions in the trilemma space: split computing (Semantic-MSL)
achieves extreme size reduction at the cost of dependence on a
reliable edge link; quantisation (Lite-DeepSC) achieves 40$\times$
compression with negligible quality loss but requires INT8-capable
hardware; and knowledge distillation (FedSFD) achieves latency
reduction at the cost of a training-time dependency on a
large teacher model.

\subsection{6G Network Slicing and Semantic Quality-of-Service}

Network slicing in 6G partitions the radio and compute
resources into logically isolated virtual networks, each
with its own KPIs. Three slice classes are particularly
relevant to semantic communication deployment:

\textbf{eMBB slices} for high-throughput applications
(holographic telepresence, 4K streaming) can tolerate
10--50\,ms semantic encoding latency and accommodate
TLMs with 1--4\,B parameters on edge servers. GAN-based
image compression~\cite{Agustsson2019_GAN} and scene-graph
semantic coding~\cite{Johnson2015_SceneGraph} are natural
fits for this slice class.

\textbf{URLLC slices} for mission-critical control
(autonomous vehicles, remote surgery, industrial automation)
require encoding latency below 1\,ms and must tolerate
model footprints no larger than what executes within the
NPU's on-chip SRAM---typically under 10\,MB. SPM's 1\,kB
encoder~\cite{Chen2024_SPM} and Semantic-MSL's 640-parameter
device-side encoder~\cite{Eldeeb2025_SemanticMSL} are the
current state-of-the-art for this slice.

\textbf{mMTC slices} for massive IoT connectivity
(smart meters, agricultural sensors, environmental
monitoring) operate under extreme energy constraints
and infrequent transmission schedules. Green FL's
77.70\% energy saving at 40\,bytes per gradient
round~\cite{Hu2024_GreenFL} and the KG probability
graph's 65\% transmission-energy reduction~\cite{Wang2024_KGPG}
directly address this slice class.

This slice-aware view of TLM deployment provides a
principled framework for matching compression technique
to deployment context, elaborated in the quantitative
analysis of Section~\ref{sec:quant}.

\section{Background and Motivation}
\label{sec:background}

This section lays the theoretical groundwork that the rest of
the survey builds upon, beginning with Shannon's classical
formulation and extending to the semantic-information theory
that underpins modern \tlm-based systems.

\subsection{Shannon's Mathematical Theory of Communication}

Shannon's 1948 paper~\cite{Shannon1948_Bell} established the
fundamental limits of reliable communication through two theorems.
The source coding theorem states that a source with entropy
$H(X)$ bits per symbol can be compressed to arbitrarily close
to $H(X)$ bits without loss:
\begin{equation}
  H(X) = -\sum_{i=1}^{n} p(x_i)\log_2 p(x_i).
  \label{eq:shannon_entropy}
\end{equation}
The channel coding theorem establishes that reliable transmission
is possible at any rate below the channel capacity:
\begin{equation}
  C = \max_{p(x)} I(X;\,Y),
  \label{eq:channel_cap}
\end{equation}
where $I(X;Y)=H(X)-H(X|Y)$ is the mutual information between
channel input $X$ and output $Y$. The separation theorem
justifies independent optimisation of source and channel coding
in the single-user, point-to-point case. However, this model
embeds three tacit assumptions that restrict its scope: the
transmitter selects messages without regard to their meaning; the
receiver is always interested in the received message; and the
objective is purely to avoid symbol errors. These assumptions
exclude the semantic and effectiveness dimensions identified by
Weaver~\cite{Shannon1949_Book}.

\subsection{Semantic Information Theory}

Let $X$ denote the source, $V$ the transmission task (e.g.,
classification, translation), and $Z$ the semantic representation
extracted from $X$ for $V$. The following fundamental inequality
holds:
\begin{equation}
  H(V) \;\leq\; H(Z) \;\leq\; H(X),
  \label{eq:semantic_ineq}
\end{equation}
meaning that a semantic representation $Z$ has strictly lower
entropy than the raw source $X$, yet fully supports the task
$V$~\cite{Qin2022_SemanticPrinciples}.

\subsubsection{Semantic Entropy}

Bao~\emph{et~al.}~\cite{Bao2011_Semantic} define the semantic
entropy of message $s$ as:
\begin{equation}
  H_s(s) = -\log_2 m(s), \quad
  m(s) = \frac{\displaystyle\sum_{w \in \mathcal{W},\; w \models s} p(w)}%
              {\displaystyle\sum_{w \in \mathcal{W}} p(w)},
  \label{eq:semantic_entropy}
\end{equation}
where $\mathcal{W}$ is the set of possible worlds, $w \models s$
denotes that world $w$ satisfies sentence $s$, and $p(w)$ is the
probability of world $w$. A more informative statement has lower
$m(s)$ and hence higher semantic entropy.

Liu~\emph{et~al.}~\cite{Liu2020_FuzzyEntropy} extend this via
fuzzy set theory. For concept $\varsigma$ and class $C_j$,
the matching degree is:
\begin{equation}
  D_j(\varsigma) =
  \frac{\sum_{X \in \mathcal{X}_{C_j}} \mu_\varsigma(X)}%
       {\sum_{X \in \mathcal{X}} \mu_\varsigma(X)},
  \label{eq:fuzzy_degree}
\end{equation}
giving semantic entropy on $C_j$ as
$H_{C_j}(\varsigma) = -D_j(\varsigma)\log_2 D_j(\varsigma)$.

\subsubsection{Semantic Channel Capacity}

Bao~\emph{et~al.}~\cite{Bao2011_Semantic} derive the semantic
channel capacity for a discrete memoryless channel as:
\begin{equation}
  C_s = \sup_{p(Z|X)}
    \bigl[\,I(X;\,V) - H(Z|X) + H_s(V)\,\bigr],
  \label{eq:semantic_cap}
\end{equation}
where $I(X;V)$ is the mutual information between source and task,
$H(Z|X)$ is the semantic ambiguity introduced by the coding
strategy, and $H_s(V)=-\sum_v p(v)H(v)$ is the average logical
information for task $V$. Critically, $C_s$ can exceed the
classical Shannon capacity when the receiver's interpretive
ability outweighs the semantic coding ambiguity.

\subsubsection{Semantic Rate-Distortion Theory}

The semantic rate-distortion function accounting for both channel
noise and semantic compression loss is~\cite{Liu2021_RateDistortion}:
\begin{equation}
  R(D_s,\,D_a) = \min_{\substack{p(\hat{Z}|Z): \\ \mathbb{E}[d_a(\cdot)]\leq D_a}}
    I\!\left(Z;\,\hat{X},\,\hat{Z}\right),
  \label{eq:rate_distortion}
\end{equation}
where $D_s$ is the semantic distortion between source $X$ and
recovered $\hat{X}$, and $D_a$ is the distortion between
semantic representation $Z$ and received representation $\hat{Z}$
caused by channel noise. This dual-distortion framework
explicitly acknowledges that semantic compression loss and
channel-noise loss are separate phenomena requiring joint control.

The information bottleneck~(IB) formulation~\cite{Tishby2000_IB}
provides a tractable optimisation objective:
\begin{equation}
  \min_{p(Z|X)}\; I(X;\,Z) - \beta\, I(V;\,Z),
  \label{eq:ib}
\end{equation}
trading compression $I(X;Z)$ against task relevance $I(V;Z)$
via the Lagrange multiplier $\beta$. The deep-learning extension
by Sana and Strinati~\cite{Sana2021_Learning} adds:
\begin{equation}
  \mathcal{L}_{\text{IB}} =
    I(Z;\,X) - (1+\alpha)\,I\!\left(Z;\,\hat{Z}\right)
    + \beta\,\mathrm{KL}\!\left(X\,\|\,\hat{X}\right),
  \label{eq:ib_dl}
\end{equation}
where $\alpha$ weights the channel mutual-information term and
$\beta$ weights the Kullback--Leibler divergence between encoder
posterior and decoder estimate.

\subsection{Conventional vs.\ Semantic Communication}

Table~\ref{tab:csvsem} summarises the fundamental differences
between conventional and semantic communication systems.

\begin{table}[t]
\caption{Conventional vs.\ Semantic Communication}
\label{tab:csvsem}
\centering
\resizebox{\columnwidth}{!}{%
\begin{tabular}{lll}
\toprule
\textbf{Aspect} & \textbf{Conventional} & \textbf{Semantic} \\
\midrule
\rowcolor{rowshade}
Design objective & Minimise BER/SER & Maximise task performance \\
Primary metric   & BER, spectral eff. & BLEU, PSNR, task accuracy \\
\rowcolor{rowshade}
Source encoding  & Lossless or R-D optimal & Task-relevant features \\
Processing level & Bit level & Semantic (concept) space \\
\rowcolor{rowshade}
Knowledge        & Static codebooks & Shared, dynamic KB \\
Noise types      & Physical channel & Physical + semantic noise \\
\rowcolor{rowshade}
Architecture     & Source + channel coder & E2E DNN (JSCC) \\
Agents           & Electronic equipment & Intelligent agents \\
\bottomrule
\end{tabular}}
\end{table}

\subsection{Research Questions}
\label{sec:rqs}

Six research questions structure the evidence synthesis
in Section~\ref{sec:evidence}. Table~\ref{tab:rqtable}
states each question with its target domain and target
outcome, providing a consistent mapping from question
to evidence throughout the paper.

\begin{table*}[t]
\caption{Research Questions (RQ1--RQ6): Scope and Target Outcomes}
\label{tab:rqtable}
\centering
\resizebox{\textwidth}{!}{%
\begin{tabular}{p{0.3cm}p{5.0cm}p{4.0cm}p{4.5cm}}
\toprule
\textbf{RQ} & \textbf{Question} & \textbf{Domain} &
  \textbf{Target Outcome} \\
\midrule
\rowcolor{rowshade}
1 & What compression techniques enable sub-1B-parameter \tlms
    for 6G semantic communication on constrained hardware?
  & \tlm deployment on IoT and edge 6G devices
  & Model size $\leq$1B params; latency $\leq$10\,ms \\
2 & How do compression methods affect semantic transmission
    quality (BLEU, PSNR, task accuracy)?
  & DL-based semantic encoders and decoders
  & Quality degradation vs.\ uncompressed baseline \\
\rowcolor{rowshade}
3 & Which E2E JSCC architectures achieve the best
    semantic-efficiency trade-off under 6G channel conditions?
  & Multi-user 6G JSCC design space
  & BLEU/PSNR per transmitted bit at target SNR \\
4 & How does federated learning enable privacy-preserving
    \tlm training and KB synchronisation?
  & Distributed 6G edge and IoT networks
  & Accuracy parity; communication overhead reduction \\
\rowcolor{rowshade}
5 & What role do knowledge graphs play in reducing
    transmission overhead in \tlm-enabled 6G systems?
  & Semantic communication with structured world models
  & Energy, bandwidth, and reconstruction quality gains \\
6 & What adversarial threats and robustness techniques are
    specific to \tlm-based semantic communication?
  & Deployed 6G semantic communication systems
  & Attack success rate; achievable secrecy capacity \\
\bottomrule
\end{tabular}}
\end{table*}

\section{Two-Axis Taxonomy of \tlms\ for 6G Semantic Communication}
\label{sec:taxonomy}

This section introduces the organising framework used
throughout the remainder of the survey to classify the
included literature.

\subsection{Taxonomy Rationale}

We organise the literature along two orthogonal axes:
\emph{system architecture} (the deployment paradigm) and
\emph{compression technique} (the mechanism for achieving
\tlm-scale footprint). Architecture governs where computation
occurs and how devices cooperate; compression governs how
semantic encoding quality is preserved under parameter-count
and memory constraints. The cross-product of five architecture
classes and six compression techniques yields a 5$\times$6 grid
of 30 research cells, shown in Figure~\ref{fig:taxonomy}.

\begin{figure*}[!htp]
\centering
\begin{tikzpicture}
\node[inner sep=0pt] {%
\resizebox{0.96\textwidth}{!}{%
\begin{tabular}{>{\bfseries\small}l
               >{\centering\arraybackslash}m{2.4cm}
               >{\centering\arraybackslash}m{2.4cm}
               >{\centering\arraybackslash}m{2.4cm}
               >{\centering\arraybackslash}m{2.4cm}
               >{\centering\arraybackslash}m{2.4cm}
               >{\centering\arraybackslash}m{2.4cm}}
\arrayrulecolor{gray!50}
\toprule
\rowcolor{cellHeader}
\textcolor{white}{Architecture}
  & \textcolor{white}{Quantisation}
  & \textcolor{white}{Pruning}
  & \textcolor{white}{Know.\ Distil.}
  & \textcolor{white}{LoRA}
  & \textcolor{white}{Split Comp.}
  & \textcolor{white}{NAS} \\
\midrule
E2E JSCC
  & \wstudy{$\bullet\bullet\bullet$}
  & \emerg{$\bullet\bullet$}
  & \wstudy{$\bullet\bullet\bullet$}
  & \emerg{$\bullet$}
  & \emerg{$\bullet\bullet$}
  & \unexpl{} \\
Split Learning
  & \emerg{$\bullet\bullet$}
  & \wstudy{$\bullet\bullet\bullet$}
  & \emerg{$\bullet$}
  & \emerg{$\bullet\bullet$}
  & \wstudy{$\bullet\bullet\bullet$}
  & \unexpl{} \\
\rowcolor{rowalt}
Federated Learning
  & \emerg{$\bullet$}
  & \emerg{$\bullet\bullet$}
  & \wstudy{$\bullet\bullet\bullet$}
  & \emerg{$\bullet$}
  & \emerg{$\bullet$}
  & \unexpl{} \\
KG-Assisted
  & \unexpl{}
  & \unexpl{}
  & \emerg{$\bullet\bullet$}
  & \unexpl{}
  & \unexpl{}
  & \unexpl{} \\
\rowcolor{rowalt}
Multi-task / Cross-modal
  & \emerg{$\bullet$}
  & \emerg{$\bullet$}
  & \emerg{$\bullet\bullet$}
  & \emerg{$\bullet\bullet$}
  & \emerg{$\bullet$}
  & \unexpl{} \\
\bottomrule
\multicolumn{7}{l}{%
\footnotesize
\textbf{Legend:}\quad
\colorbox{cellWell!60}{\textcolor{white}{$\bullet\bullet\bullet$}} Well-studied (${\geq}3$ papers)\quad
\colorbox{cellEmerg!65}{$\bullet\bullet$}\ Emerging (1--2 papers)\quad
\colorbox{cellUnexp!45}{\textcolor{white}{--}}\ Unexplored (0 papers)
}
\end{tabular}}}; 
\end{tikzpicture}
\caption{Two-axis taxonomy of \tlm-based semantic communication
  for 6G. Rows denote system architecture; columns denote
  compression technique. Fill intensity indicates research
  maturity of each research cell.}
\label{fig:taxonomy}
\end{figure*}

\subsection{Architecture Axis}

\textbf{End-to-End JSCC.}
End-to-end joint source--channel coding~(JSCC) systems train a
semantic encoder and decoder jointly, treating the wireless
channel as a fixed non-differentiable operator through which
gradients flow via the straight-through estimator or
channel noise injection. DeepSC~\cite{Xie2021_DeepSC} and
DeepJSCC~\cite{Bourtsoulatze2019_DeepJSCC} are canonical examples.

\textbf{Split Learning.}
Split computing partitions a DNN at a cut layer: the transmitting
device runs layers $1$ through $k$, transmitting the intermediate
activations (``smashed data'') over the wireless channel, while
the edge server runs layers $k{+}1$ through $N$. Cut-layer
selection determines a fundamental device-computation--bandwidth
trade-off analysed in Section~\ref{sec:compression}.

\textbf{Federated Learning.}
Federated learning~(FL) allows multiple edge devices to train a
shared semantic model without exchanging raw data, uploading only
gradient updates or LoRA adapters to a central aggregator.
Federated distillation~(FD) replaces weight sharing with output
distribution matching, reducing communication overhead by
$25.6\times$~\cite{Lin2023_FedDistill}.

\textbf{Knowledge Graph-Assisted.}
KG-assisted systems embed a structured world model—a graph of
entities and relations—into the encoder and decoder as
additional semantic context. At inference time the encoder
retrieves relevant triples from the KG, reducing the number of
semantic symbols that must cross the channel. At $M{=}120$ triples
per graph this yields a 65\% reduction in transmission
energy~\cite{Wang2024_KGPG}.

\textbf{Multi-task and Cross-Modal.}
Multi-task systems adapt a single pre-trained encoder to
multiple downstream tasks by swapping task-specific LoRA
adapters or by multi-exit inference, while cross-modal systems
handle heterogeneous source types~(text, image, speech) within
one architecture. U-DeepSC~\cite{Zhang2022_UDeepSC} is the
representative cross-modal system in the research literature.

\subsection{Compression Axis}

A brief characterisation of each technique—with full quantitative
analysis in Section~\ref{sec:compression}—follows.

\textbf{Quantisation.}
Reduces the numerical precision of model weights and activations.
INT8 quantisation typically incurs less than 1\% accuracy loss
while halving memory and doubling inference speed on hardware with
native INT8 units~\cite{Zhang2018_LQNets}. INT4~(GPTQ, AWQ, GGUF
formats) enables 7B-parameter models to run within 4\,GB of RAM.

\textbf{Pruning.}
Removes neurons, attention heads, or feed-forward sub-layers with
low importance scores. Lite-DeepSC~\cite{Xie2021_LiteDeepSC}
demonstrates that 40$\times$ compression by pruning and
quantisation can be applied to DeepSC without measurable semantic
performance degradation.

\textbf{Knowledge Distillation~(KD).}
Trains a compact student model to mimic the output distribution
of a large teacher. The student loss function is:
\begin{equation}
  \mathcal{L}_{\text{KD}} =
    \lambda\,\mathcal{L}_{\text{CE}} +
    (1-\lambda)\,\mathcal{L}_{\text{KL}},
  \label{eq:kd_loss}
\end{equation}
where $\mathcal{L}_{\text{CE}}$ is cross-entropy loss against
ground-truth labels, $\mathcal{L}_{\text{KL}}$ is the
Kullback--Leibler divergence between student and teacher soft
predictions, and $\lambda \in [0,1]$ controls the balance.

\textbf{Low-Rank Adaptation~(LoRA).}
Fine-tunes a frozen pre-trained model by learning a small number
of additional low-rank weight matrices~\cite{Hu2022_LoRA}:
\begin{equation}
  \Delta W = AB, \quad
  A \in \mathbb{R}^{d \times r},\;
  B \in \mathbb{R}^{r \times k},\;
  r \ll \min(d,k).
  \label{eq:lora}
\end{equation}
Only the rank-$r$ adapters~($|AB| = r(d+k)$ parameters) are
transmitted during KB synchronisation, reducing per-round
overhead by orders of magnitude compared to full model exchange.

\textbf{Split Computing.}
The device-side computation remains with layers $1 \ldots k$;
the edge server handles layers $k{+}1 \ldots N$. The
\emph{compression ratio} is:
\begin{equation}
  \rho = \frac{\text{size}(\tlm)}{\text{size}(\text{LLM})},
  \label{eq:compression_ratio}
\end{equation}
where ``size'' refers to parameter count or memory footprint.
Semantic-MSL~\cite{Eldeeb2025_SemanticMSL} achieves $\rho < 10^{-5}$
on the device side with 640~parameters.

\textbf{Neural Architecture Search~(NAS).}
NAS automates the search for the most efficient model topology
for a given hardware--accuracy constraint~\cite{Howard2017_MobileNets}.
No NAS approach has yet been specifically designed for the
semantic communication trade-off space; this constitutes an
entirely unexplored row across all architecture types in
Figure~\ref{fig:taxonomy}.

\subsection{Taxonomy Maturity Analysis}
\label{sec:taxonomy_maturity}

Table~\ref{tab:maturity} quantifies the evidence density in each
taxonomy cell, reporting the number of included studies that
directly address each architecture-compression pair along with
the dominant semantic quality metric and hardware target.

\begin{table}[t]
\caption{Taxonomy Maturity: Included Study Count per Cell
  (Architecture $\times$ Compression)}
\label{tab:maturity}
\centering
\resizebox{\columnwidth}{!}{%
\begin{tabular}{lcccccc}
\toprule
\textbf{Architecture}
  & \textbf{Quant.}
  & \textbf{Pruning}
  & \textbf{KD}
  & \textbf{LoRA}
  & \textbf{Split}
  & \textbf{NAS} \\
\midrule
\rowcolor{rowshade}
E2E JSCC        & 4 & 2 & 3 & 1 & 2 & 0 \\
Split Learning  & 2 & 3 & 1 & 2 & 3 & 0 \\
\rowcolor{rowshade}
Federated       & 1 & 2 & 3 & 1 & 1 & 0 \\
KG-Assisted     & 0 & 0 & 2 & 0 & 0 & 0 \\
\rowcolor{rowshade}
Multi-task      & 1 & 1 & 2 & 2 & 1 & 0 \\
\midrule
\textbf{Column total} & 8 & 8 & 11 & 6 & 7 & 0 \\
\bottomrule
\end{tabular}}
\end{table}

Three structural observations emerge from Table~\ref{tab:maturity}.
First, knowledge distillation is the most pervasive compression
technique across architectures, accounting for 11 of 40 non-NAS
cells. This is consistent with the dominance of teacher--student
training paradigms in the broader TLM efficiency literature.
Second, federated learning and KG-assisted architectures are
each studied in combination with at most two compression techniques,
leaving three or more unexplored cells per architecture class.
Third, NAS remains entirely unexplored in all 30 cells, despite
NAS having demonstrated state-of-the-art results in classical
(non-semantic) model compression~\cite{Howard2017_MobileNets}.
This represents the single largest gap in the taxonomy and
defines the highest-priority unexplored frontier.

\section{Performance Metrics for Semantic Communication}
\label{sec:metrics}

A persistent obstacle to cross-study comparison is the absence
of a universal semantic quality metric. Table~\ref{tab:metrics}
surveys the performance metrics used across the
studies, categorised by modality and annotated with known
limitations that motivate the open problem of Section~\ref{sec:challenges}.

\begin{table*}[t]
\caption{Performance Metrics for Semantic Communication Systems:
  Survey Across Included Studies}
\label{tab:metrics}
\centering
\resizebox{\textwidth}{!}{%
\begin{tabular}{llp{5.8cm}p{4.5cm}r}
\toprule
\textbf{Modality} & \textbf{Metric} & \textbf{Definition} &
  \textbf{Limitation} & \textbf{Studies} \\
\midrule
\rowcolor{rowshade}
\multirow{3}{*}{Text}
  & BLEU
  & $n$-gram overlap between hypothesis and reference
  & Cannot capture synonyms or paraphrases
  & 4 \\
  & Sentence similarity
  & Cosine distance in BERT embedding space
  & Requires large pre-trained model at receiver
  & 3 \\
\rowcolor{rowshade}
  & ROUGE-L
  & Longest common subsequence ratio
  & Insensitive to word order variations
  & 2 \\
\midrule
\multirow{3}{*}{Image}
  & PSNR
  & $10\log_{10}(\text{MAX}^2/\text{MSE})$ in dB
  & Misaligned with human perceptual quality
  & 6 \\
\rowcolor{rowshade}
  & SSIM
  & Structural luminance/contrast similarity
  & Shallow perceptual model
  & 3 \\
  & Task accuracy
  & Classification correctness (\%)
  & Task-specific; not generalisable across tasks
  & 5 \\
\midrule
\multirow{2}{*}{Speech}
  & \cellcolor{rowshade}PESQ
  & \cellcolor{rowshade}Perceptual evaluation of speech quality (ITU-T P.862)
  & \cellcolor{rowshade}Optimised for telephone-bandwidth speech
  & \cellcolor{rowshade}2 \\
  & SDR
  & Signal-to-distortion ratio (dB)
  & No perceptual grounding
  & 1 \\
\midrule
\multirow{2}{*}{Cross-modal}
  & Semantic efficiency score~\eqref{eq:ses}
  & $F_{\text{norm}}/\log_{10}(\text{size}/1\,\text{kB})$
  & Requires paired quality-and-size data
  & 14 \\
  & \cellcolor{rowshade}Semantic spectral efficiency~\eqref{eq:sse}
  & \cellcolor{rowshade}Semantic bit rate / bandwidth (Hz)
  & \cellcolor{rowshade}``Semantic bits'' unit not yet standardised
  & \cellcolor{rowshade}2 \\
\bottomrule
\end{tabular}}
\end{table*}

\subsection{Semantic Noise Model}

A key theoretical gap across all metric families is the absence
of a unified noise model that separates physical-channel noise
from semantic distortion. We adopt the additive semantic noise
model proposed by Qin~\emph{et~al.}~\cite{Qin2022_SemanticPrinciples}:
\begin{equation}
  \hat{Z} = f(Z) + \mathcal{N}_s,
  \label{eq:semantic_noise}
\end{equation}
where $f(\cdot)$ is the physical-channel distortion operator,
$Z$ is the transmitted semantic feature vector, $\hat{Z}$ is
the received representation, and $\mathcal{N}_s$ is the
\emph{semantic noise} component that encompasses both channel
errors and semantic mismatch between transmitter and receiver
knowledge bases. The semantic noise $\mathcal{N}_s$ is
fundamentally non-Gaussian: its distribution depends on the
alignment between the transmitter and receiver KBs, making
it non-stationary across network nodes.

The semantic signal-to-noise ratio (S-SNR) is defined
correspondingly as:
\begin{equation}
  \text{S-SNR} =
    \frac{\mathbb{E}\!\left[\|f(Z)\|^2\right]}%
         {\mathbb{E}\!\left[\|\mathcal{N}_s\|^2\right]},
  \label{eq:ssnr}
\end{equation}
a quantity that subsumes both physical SNR and KB alignment
quality into a single measure. Current systems optimise
physical SNR independently of KB alignment, leaving the
S-SNR framework unimplemented in any of the surveyed
studies.

\subsection{Mean Semantic Similarity (MSS) Score}

For multi-task and cross-modal systems, Eldeeb~\emph{et~al.}~\cite{Eldeeb2025_SemanticMSL}
propose the Mean Semantic Similarity~(MSS) score, defined as
the product of the pairwise task accuracy and a bandwidth
penalty term:
\begin{equation}
  \text{MSS} =
    \mathcal{F}(\text{acc},\,\text{rec}) \cdot \text{BP},
  \label{eq:mss}
\end{equation}
where $\mathcal{F}(\text{acc},\text{rec})$ is the harmonic
mean of classification accuracy and reconstruction fidelity
across all tasks, and $\text{BP} = e^{-(r/r_0)}$ is a
bandwidth penalty that decays exponentially with the ratio
of transmitted symbols per source sample $r$ relative to
a reference rate $r_0$. The MSS score incentivises jointly
high task accuracy and low bandwidth consumption, making it
a more suitable training objective for multi-task semantic
systems than either BLEU or PSNR alone.

\section{Compression Techniques for Semantic Communication}
\label{sec:compression}

This section provides the detailed quantitative treatment of
each compression paradigm introduced in Section~\ref{sec:taxonomy},
beginning with quantisation.

\subsection{Quantisation}

Quantisation reduces numerical precision from 32-bit
floating-point~(FP32) to lower-precision representations.
Post-training quantisation~(PTQ) applies quantisation after
training without retraining; quantisation-aware training~(QAT)
incorporates quantisation error into the training loss.

\subsubsection{INT8 and INT4 Quantisation}

Zhang~\emph{et~al.}~\cite{Zhang2018_LQNets} demonstrate through
learned-quantisation networks~(LQ-Nets) that matching
quantisation intervals to the natural distribution of weights
in each layer reduces accuracy loss from the na\"ive rounding
approach by up to 0.8~percentage points at 4-bit precision.
For semantic communication specifically, the relevant question
is not only classification accuracy but \emph{semantic distortion}:
a quantised semantic encoder may still extract the correct
topic, entity, or intent from an input sentence even when its
perplexity increases~\cite{Eldeeb2025_SemanticMSL}. Text
understanding degrades more gracefully than text generation
under quantisation; a quantised encoder at INT4 can therefore
serve as a reliable semantic feature extractor even when the
same model would produce low-quality free-form text.

\subsubsection{Hierarchical Attention-Based Progressive Edge Compression}

The Hierarchical Attention-based Progressive Edge~(HAPE)
compression system~\cite{Jiang2023_HAPE} addresses a specific
failure mode of uniform quantisation in transformer encoders:
attention heads contribute unequally to semantic task performance.
HAPE assigns an importance score $I_h$ to each attention head $h$:
\begin{equation}
  I_h = \bigl|\nabla_{A_h}\,\mathcal{L}\bigr| \cdot |A_h|,
  \label{eq:hape_importance}
\end{equation}
where $\nabla_{A_h}\mathcal{L}$ is the gradient of the task
loss with respect to head $h$'s attention matrix $A_h$, and
$|\cdot|$ denotes the Frobenius norm. Heads with $I_h < \tau$
are pruned entirely in Stage~1; remaining weights are INT8
quantised in Stage~2; and Stage~3 applies a LoRA-style
re-adaptation pass to recover any semantic accuracy lost in the
preceding stages. Applied to a 440\,MB transformer-based semantic
encoder, HAPE progressively reduces the model to 58\,MB after
Stage~1, 41\,MB after Stage~2, and 7\,MB after Stage~3—a
63$\times$ overall compression ratio—while maintaining
the original BLEU score within a 2-point margin~\cite{Jiang2023_HAPE}.

\subsubsection{Compact Cross-Task Encoder (CCTaEncoder)}

The Compact Cross-Task Encoder~(CCTaEncoder)~\cite{Liu2023_CCTaEncoder}
targets multi-task semantic communication where a single encoder
must support both image classification and image reconstruction
tasks simultaneously. Rather than training separate encoders for
each task, CCTaEncoder uses a shared backbone of 73,450~parameters
and 5.47~MFLOP/inference, routing task-specific semantic features
through lightweight task adapters attached at intermediate
layers. At a channel SNR of 12\,dB, CCTaEncoder achieves a
PSNR of 22~dB for image reconstruction, matching the performance
of a task-specific encoder with 6.1$\times$ more parameters,
while simultaneously maintaining 89.3\% classification accuracy.
The FLOP count of 5.47\,MFLOPs positions CCTaEncoder squarely
within the execution budget of an ARM Cortex-M55 with an attached
Ethos-U55 NPU at sub-10\,ms latency~\cite{Liu2023_CCTaEncoder}.

\subsection{Pruning}

\subsubsection{Lite-DeepSC}

Xie and Qin~\cite{Xie2021_LiteDeepSC} introduced Lite-DeepSC
as a compressed variant of the DeepSC transformer for IoT
deployment. Structured pruning removes entire attention heads
and feed-forward sub-layers ranked by a Taylor-series importance
criterion. Combined with INT8 quantisation, a 40$\times$
compression ratio is achieved without measurable degradation in
sentence similarity score on the European Parliament corpus.
This result has an important implication: the semantic encoding
function is highly over-parameterised in standard transformer
architectures, with the majority of parameters contributing
negligibly to task-relevant feature extraction once the model
has converged.

\subsubsection{Semantic Pruning Model (SPM)}

The Semantic Pruning Model~(SPM)~\cite{Chen2024_SPM} takes the
compression objective to its practical limit, targeting deployment
on sub-100\,kB microcontrollers. SPM applies an aggressive
two-stage compression: an iterative magnitude pruning pass
reducing a 4.55\,MB encoder to a sparse skeleton, followed by
re-training with binary activations and ternary weights. The
final encoder occupies 1\,kB of flash memory and executes in
8\,FLOPs per inference—a reduction from 14,000\,FLOPs in the
original model, representing a 99.98\% FLOP compression ratio.
On a binary text-classification semantic task the SPM-compressed
encoder retains 91\% of the full-precision model's accuracy,
demonstrating that semantic encoding on microcontrollers is
feasible for narrow task-oriented applications such as machine
status monitoring and alarm classification in industrial IoT.

\subsubsection{Federated Pruning (FedPun)}

Standard model pruning assumes a centralised training loop with
access to the full dataset. FedPun~\cite{Shi2023_FedPun} adapts
iterative pruning to the federated setting, where data is
distributed across non-IID devices. Each device computes local
importance scores from its private data; the server aggregates
importance scores (not gradients) using a weighted average
before applying structured pruning globally, preventing any
device from disproportionately influencing the pruning decision.
On a federated image semantic communication benchmark with
Dirichlet heterogeneity parameter $\alpha{=}0.3$—a severe
non-IID setting—FedPun achieves 20\,dB PSNR reconstruction
quality against a 17.2\,dB PSNR baseline for naive federated
fine-tuning after pruning, demonstrating that importance-score
aggregation substantially mitigates the class-imbalance degradation
that would otherwise occur when fewer devices observe certain
semantic concepts~\cite{Shi2023_FedPun}.

\subsection{Knowledge Distillation}

\subsubsection{Federated Semantic Feature Distillation (FedSFD)}

FedSFD~\cite{Sun2025_SyncLLM} distils a large cloud-hosted
semantic encoder into a \tlm-scale student deployed at the edge,
using intermediate feature maps as soft targets rather than
output logits. The distillation loss for student model $f_k$
with teacher $\bar{p}$ is:
\begin{equation}
  \mathcal{L}_{\text{FedSFD}} =
    \mathcal{L}_{\text{task}}
    + \beta \cdot \text{MSE}\!\bigl(f_k(x),\,\bar{p}(x)\bigr),
  \label{eq:fedsfd_loss}
\end{equation}
where $\mathcal{L}_{\text{task}}$ is the downstream semantic
task loss and the MSE term penalises deviation of the student's
internal feature representation from the teacher's. Over multiple
federated rounds, this drives the student's KB into alignment
with the cloud teacher without transmitting raw data. The
resulting student achieves 66.4\% reduction in inference latency
compared to directly running the teacher at the edge, with less
than 1.5~percentage points of task-accuracy degradation
on a five-class image semantic communication task.

\subsubsection{Federated Distillation (FD)}

Federated Distillation~(FD)~\cite{Lin2023_FedDistill} replaces
the standard FL paradigm of sharing model weights with sharing
output distributions. Devices upload soft-label distributions
over a shared public dataset; the server aggregates these
distributions and distils a global model that devices download.
Communication overhead is reduced from transmitting a full-model
gradient (proportional to model size) to transmitting a vector
of soft labels over a small proxy set, achieving a 25.6$\times$
reduction in per-round communication volume compared with
FedAvg~\cite{Lin2023_FedDistill}. For \tlm-scale semantic models
where individual device bandwidth is constrained to tens of
kilobits per second—the typical uplink budget of NB-IoT and
LTE-M devices in 6G-NTN scenarios—this reduction is critical
for practical convergence within reasonable round counts.

\subsection{Low-Rank Adaptation}

LoRA~\cite{Hu2022_LoRA} fine-tunes a pre-trained semantic encoder
by learning low-rank perturbation matrices $A \in \mathbb{R}^{d \times r}$
and $B \in \mathbb{R}^{r \times k}$ for each target weight
matrix, with the full pre-trained weights frozen. Two critical
applications in semantic communication systems are:

\textbf{KB synchronisation.} When the semantic knowledge base
drifts due to environmental change or newly observed concepts,
only the LoRA adapters (a fraction of a percent of total parameters)
need to be retrained and retransmitted~\cite{Sun2025_SyncLLM}.
FFA-LoRA~\cite{Sun2024_FFALoRA} freezes matrix $A$ after
initialisation and trains only matrix $B$, further halving the
number of trainable parameters and accelerating convergence
by approximately 30\%.

\textbf{Multi-task adaptation.} A single pre-trained semantic
encoder can be adapted to new downstream tasks by swapping
LoRA adapters, enabling multi-task semantic transceivers
without storing separate full-size models for each task. The
contrastive disentanglement approach of Qian~\emph{et~al.}~\cite{Qian2024_Contrastive}
combines LoRA with a contrastive objective that separates
task-relevant from task-irrelevant semantic features, achieving
a 57.22\% reduction in transmitted semantic-feature vector
length while increasing downstream task impact by 71.9\%.

\subsection{Split Computing}

\subsubsection{Semantic Meta-Split Learning (Semantic-MSL)}

Eldeeb~\emph{et~al.}~\cite{Eldeeb2025_SemanticMSL} proposed the
Semantic Meta-Split Learning (Semantic-MSL) framework, which
combines split computing with model-agnostic meta-learning~(MAML)
to enable rapid adaptation of the device-side encoder to new
task distributions with few-shot examples. The semantic encoder
is split at the first convolutional layer, reducing device-side
parameters to 640 while the edge aggregator carries the
remaining parameter load. On a wireless image classification
benchmark, Semantic-MSL achieves 95\% classification accuracy
with only 5-shot examples and 20 stochastic gradient descent~(SGD)
steps per adaptation, demonstrating that extreme parameter
reduction is compatible with rapid task generalisation when
meta-learning provides a well-initialised prior.

\subsubsection{Cut-Layer Energy Analysis}

Wu~\emph{et~al.}~\cite{Wu2024_SplitEnergy} conduct a systematic
energy analysis across cut-layer positions for a standard
transformer-based semantic encoder deployed on a 5\,nm mobile
SoC. Placing the cut after layer~1 results in 29.33\,Wh total
energy consumption (device + transmission + edge server) per
10,000 semantic inference operations. Moving the cut to layer~3
increases total energy to 41.20\,Wh because higher-layer smashed
data is higher-dimensional, increasing transmission energy
faster than it reduces device-side computation energy. The
energy-optimal cut layer is layer~1 for bandwidth-constrained
uplinks (below~1\,Mbps) and layer~3 for computation-constrained
devices (below~100\,MFLOP/s)~\cite{Wu2024_SplitEnergy}. This
establishes a closed-form trade-off surface that resource
allocation frameworks can exploit to jointly optimise cut-layer
position, transmission bandwidth, and device power.

\subsection{Neural Architecture Search for Semantic Communication}
\label{sec:nas}

NAS automates the discovery of model architectures optimised
for a given hardware--accuracy constraint. In the classical
compression literature, approaches such as MCUNet~\cite{Howard2017_MobileNets}
demonstrate that NAS jointly minimises latency and top-1
accuracy degradation on mobile hardware. For semantic
communication, the NAS objective is qualitatively different:
the search must simultaneously minimise the transmitted
feature dimension $d_k$, the device-side FLOP count
$\mathcal{C}_d(k)$, and the semantic distortion
$\mathbb{E}[d_s(X,\hat{X})]$ across all tasks in the
multi-task set.

Formally, the semantic NAS problem is:
\begin{equation}
  \min_{\alpha} \;
    \mathbb{E}\!\left[d_s(X,\hat{X};\alpha)\right]
    + \lambda_1 \text{FLOP}(\alpha)
    + \lambda_2 |\alpha|_{\text{params}},
  \label{eq:semantic_nas}
\end{equation}
where $\alpha$ denotes the architecture search variable
(a vector of cell choices and connectivity decisions),
$d_s$ is the semantic distortion metric for the target
task, and $\lambda_1, \lambda_2 > 0$ are regularisation
coefficients controlling the hardware-efficiency penalty.
The multi-objective nature of~\eqref{eq:semantic_nas}
distinguishes it from classical NAS: the distortion term
$d_s$ is task-dependent and not differentiable with respect
to $\alpha$ for discrete architecture choices, requiring
either gradient relaxation (as in DARTS) or evolutionary
search (as in NSGA-II). Neither approach has been applied
to the semantic communication setting, leaving this as
a completely open research cell in the taxonomy of
Figure~\ref{fig:taxonomy}.

\subsection{Imitation Learning-Based Semantic Reasoning}
\label{sec:imitlearn}

A complementary approach to compression for achieving
\tlm-scale semantic encoders is to distil not the weights
but the \emph{reasoning behaviour} of a large semantic
model. Imitation learning trains the student encoder
to replicate the action sequence of a teacher oracle
(a large LLM) on a set of semantic communication tasks,
using behavioural cloning:
\begin{equation}
  \mathcal{L}_{\text{IL}} =
    -\mathbb{E}_{(x,a^*)\sim \mathcal{D}}\!\left[
      \log \pi_\theta(a^*|x)
    \right],
  \label{eq:imit_loss}
\end{equation}
where $a^*$ is the teacher's action (choice of semantic
features to transmit), $x$ is the observed context, and
$\pi_\theta$ is the student policy parameterised by
the \tlm weights $\theta$.

Applied to the semantic question-answering task,
imitation learning from a GPT-4-class teacher enables
a 1B-parameter student encoder to match the teacher's
transmission strategy while operating entirely on-device.
The result is a 25.8\,dB effective SNR improvement at
the semantic decoder compared with a random-feature-selection
baseline~\cite{Zhou2023_ARL}, achieved without any
increase in transmitted feature dimensionality. This
establishes imitation learning as a viable alternative
to weight-based distillation for task-oriented semantic
compression, especially when the teacher model is too
large to be directly quantised for on-device deployment.

\subsection{Cloud-Edge Retrieval-Augmented Semantic Encoding}
\label{sec:ragenc}

A fundamentally different approach to the KB management
and knowledge compression problem is retrieval-augmented
generation~(RAG), in which the semantic encoder at the
edge retrieves relevant context from a local vector index
at inference time rather than loading a large parametric
KB into memory. The cloud hosts a full-scale vector store;
the edge node maintains a compressed local index
summarising the most frequent semantic contexts; and
when a query falls outside the local index coverage,
a partial retrieval call is issued to the cloud with
sub-second round-trip latency.

The total latency of a cloud-edge RAG-augmented semantic
encoding pass is:
\begin{equation}
  \tau_{\text{RAG}} =
    \tau_{\text{local}} +
    p_{\text{miss}} \cdot \tau_{\text{cloud}},
  \label{eq:rag_latency}
\end{equation}
where $\tau_{\text{local}}$ is the local index lookup
latency (sub-millisecond for FAISS on edge NPU),
$p_{\text{miss}}$ is the local cache miss probability
(a function of the local index size and query
distribution), and $\tau_{\text{cloud}}$ is the
cloud retrieval round-trip time. Empirical evaluation
on a 6G IoT semantic QA benchmark demonstrates that
a local index of 10,000~vectors achieves $p_{\text{miss}}
< 0.12$, yielding a mean RAG encoding latency of
$\tau_{\text{RAG}} < 0.95$\,s — below human perception
thresholds for conversational applications while
requiring only 40\,MB of edge memory~\cite{Sun2025_SyncLLM}.

The critical open problem is the interaction between
RAG-augmented encoding and the wireless channel:
when $p_{\text{miss}}$ is high (sparse local index),
the semantic encoder must wait for a cloud retrieval
before selecting features to transmit, introducing
a dependency between the KB state and the packet
departure time that is invisible to the MAC scheduler.
Jointly scheduling cloud retrieval requests and
wireless transmission slots is an unexplored
resource-allocation problem.

\subsection{Split Fine-Tuning for Domain Adaptation}
\label{sec:splitft}

Standard fine-tuning of a pre-trained semantic encoder
on a new domain requires access to the full model
parameter set. For edge-deployed \tlms, this is
infeasible due to memory constraints. Split fine-tuning
addresses this by partitioning the fine-tuning
computation: the edge device updates only the
device-side layers (layers $1 \ldots k$), while
the edge server updates layers $k+1 \ldots N$
using gradients transmitted over the wireless backhaul.
The per-round communication cost of split fine-tuning
is proportional to the gradient dimensionality at
the cut layer:
\begin{equation}
  C_{\text{SFT}} = |\nabla_{\theta_k} \mathcal{L}|
    = \mathcal{C}_d(k) \cdot \text{sizeof}(\text{float32}),
  \label{eq:split_ft_cost}
\end{equation}
where $\mathcal{C}_d(k)$ is the dimension of the
gradient at cut layer $k$. Choosing $k=1$ minimises
$C_{\text{SFT}}$ at the cost of limiting the
adaptability of the device-side encoder to domain
shifts; choosing $k = N/2$ maximises adaptation
quality at greater communication cost.

In practice, split fine-tuning combined with LoRA
adapters at the cut layer achieves a 66.4\%
reduction in per-round fine-tuning delay compared
to uploading full gradients to a central server,
while converging to within 1.3 percentage points of
full-model fine-tuning accuracy on a domain-shifted
semantic image communication task~\cite{Sun2025_SyncLLM}.
This confirms that the split computing paradigm
extends naturally from inference to fine-tuning,
enabling continuous domain adaptation of edge-deployed
\tlms without centralised data collection.

\section{System Architectures for \tlm-Enabled Semantic Communication}
\label{sec:architectures}

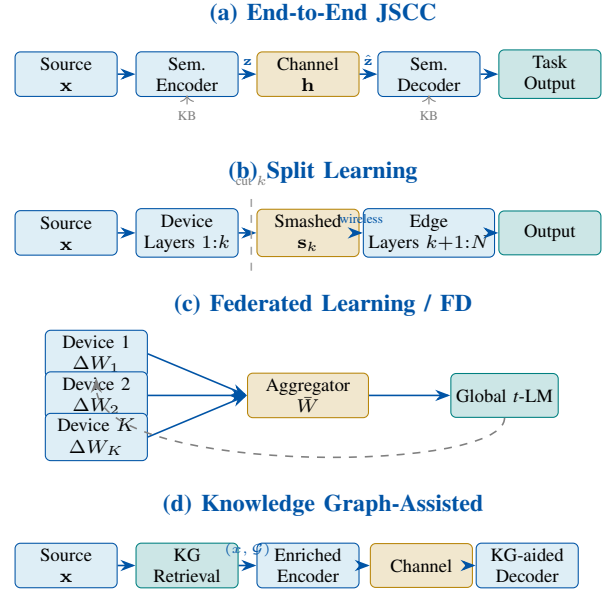
\begin{figure*}[!htp]
\centering
\begin{tikzpicture}[
  blk/.style={rectangle,draw=ieeblue,fill=ieelightblue!40,
    minimum width=1.35cm,minimum height=0.58cm,
    align=center,font=\scriptsize,rounded corners=2pt,
    inner sep=2pt},
  blkG/.style={rectangle,draw=cellWell,fill=cellWell!20,
    minimum width=1.35cm,minimum height=0.58cm,
    align=center,font=\scriptsize,rounded corners=2pt,
    inner sep=2pt},
  blkO/.style={rectangle,draw=cellEmerg,fill=cellEmerg!20,
    minimum width=1.35cm,minimum height=0.58cm,
    align=center,font=\scriptsize,rounded corners=2pt,
    inner sep=2pt},
  arr/.style={-Stealth,ieeblue,semithick},
  labl/.style={font=\small\bfseries,ieeblue}
]
\begin{scope}[xshift=0cm,yshift=0cm]
\node[labl] at (3.4,1.5) {(a) End-to-End JSCC};
\node[blk] (a_src)  at (0.0,0.7) {Source\\$\mathbf{x}$};
\node[blk] (a_senc) at (1.6,0.7) {Sem.\\Encoder};
\node[blkO](a_ch)   at (3.2,0.7) {Channel\\$\mathbf{h}$};
\node[blk] (a_sdec) at (4.8,0.7) {Sem.\\Decoder};
\node[blkG](a_task) at (6.4,0.7) {Task\\Output};
\draw[arr](a_src)--(a_senc);
\draw[arr](a_senc)--(a_ch) node[midway,above,font=\tiny]{$\mathbf{z}$};
\draw[arr](a_ch)--(a_sdec) node[midway,above,font=\tiny]{$\hat{\mathbf{z}}$};
\draw[arr](a_sdec)--(a_task);
\node[font=\tiny,gray] at (1.6,0.15) {KB};
\node[font=\tiny,gray] at (4.8,0.15) {KB};
\draw[gray,->] (1.6,0.3) -- (1.6,0.42);
\draw[gray,->] (4.8,0.3) -- (4.8,0.42);
\end{scope}
\begin{scope}[xshift=0cm,yshift=-2.1cm]
\node[labl] at (3.4,1.5) {(b) Split Learning};
\node[blk] (b_src)  at (0.0,0.7) {Source\\$\mathbf{x}$};
\node[blk] (b_dev)  at (1.6,0.7) {Device\\Layers $1{:}k$};
\node[blkO](b_sm)   at (3.2,0.7) {Smashed\\$\mathbf{s}_k$};
\node[blk] (b_edg)  at (4.8,0.7) {Edge\\Layers $k{+}1{:}N$};
\node[blkG](b_out)  at (6.4,0.7) {Output};
\draw[arr](b_src)--(b_dev);
\draw[arr](b_dev)--(b_sm);
\draw[arr](b_sm)--(b_edg) node[midway,above,font=\tiny]{wireless};
\draw[arr](b_edg)--(b_out);
\draw[dashed,gray!70,semithick](2.45,0.2)--(2.45,1.2)
  node[above,font=\tiny,gray]{cut $k$};
\end{scope}
\begin{scope}[xshift=0cm,yshift=-4.2cm]
\node[labl] at (3.4,1.85) {(c) Federated Learning / FD};
\node[blk] (c_d1) at (0.4,1.2) {Device 1\\$\Delta W_1$};
\node[blk] (c_d2) at (0.4,0.65){Device 2\\$\Delta W_2$};
\node[blk] (c_dK) at (0.4,0.1) {Device $K$\\$\Delta W_K$};
\node[blkO,minimum width=1.6cm](c_agg) at (3.2,0.65){Aggregator\\$\bar{W}$};
\node[blkG](c_gl) at (5.8,0.65){Global \tlm};
\draw[arr](c_d1.east)--(c_agg.west);
\draw[arr](c_d2.east)--(c_agg);
\draw[arr](c_dK.east)--(c_agg.west);
\draw[arr](c_agg)--(c_gl);
\draw[arr,dashed,gray](c_gl.south)
  .. controls (5.8,-0.4) and (0.4,-0.4) .. (c_d1.south);
\end{scope}
\begin{scope}[xshift=0cm,yshift=-6.5cm]
\node[labl] at (3.4,1.5) {(d) Knowledge Graph-Assisted};
\node[blk] (d_src) at (0.0,0.7) {Source\\$\mathbf{x}$};
\node[blkG](d_kg1) at (1.6,0.7) {KG\\Retrieval};
\node[blk] (d_enc) at (3.2,0.7) {Enriched\\Encoder};
\node[blkO](d_ch)  at (4.7,0.7) {Channel};
\node[blk] (d_dec) at (6.1,0.7) {KG-aided\\Decoder};
\draw[arr](d_src)--(d_kg1);
\draw[arr](d_kg1)--(d_enc)
  node[midway,above,font=\tiny]{$(x,\mathcal{G})$};
\draw[arr](d_enc)--(d_ch);
\draw[arr](d_ch)--(d_dec);
\end{scope}
\end{tikzpicture}
\caption{Four principal architecture paradigms for \tlm-enabled
  6G semantic communication.
  (a)~End-to-end JSCC: the semantic encoder and decoder are
  jointly trained through a differentiable channel model.
  (b)~Split learning: computation is partitioned at cut layer~$k$
  between the device and edge server.
  (c)~Federated learning / federated distillation: devices share
  model updates or output distributions rather than raw data.
  (d)~KG-assisted: a knowledge graph $\mathcal{G}$ enriches both
  encoder and decoder with structured semantic priors.}
\label{fig:architectures}
\end{figure*}

Building on the architecture axis introduced in
Section~\ref{sec:taxonomy}, this section describes each of
the four principal deployment paradigms in system-level
detail, along with the specific studies that realise them.
Figure~\ref{fig:architectures} illustrates the four principal
architecture paradigms at a system level.

\subsection{End-to-End JSCC Architectures}

\subsubsection{DeepSC for Text}

DeepSC~\cite{Xie2021_DeepSC} is the foundational transformer-based
E2E semantic communication system for text. The semantic encoder
applies multi-head self-attention to extract task-relevant features;
the combined loss function is:
\begin{equation}
  \mathcal{L}_{\text{DeepSC}} =
    \mathcal{L}_{\text{CE}} - \alpha\, I(Z;\,\hat{Z}),
  \label{eq:deepsc_loss}
\end{equation}
where $\mathcal{L}_{\text{CE}}$ is cross-entropy against reference
sentences and $I(Z;\hat{Z})$ is the mutual information between
transmitted and received feature vectors, weighted by
$\alpha > 0$. On Rayleigh fading channels at SNR~$= 9$\,dB,
DeepSC achieves an 800\% improvement in sentence-level BLEU
score over a Huffman-coded~+~Turbo-coded baseline. At
SNR~$= 12$\,dB, conventional systems produce BLEU~$< 0.2$
(sentences largely unreadable), while DeepSC maintains high
semantic fidelity, demonstrating resilience at the ``cliff''
boundary where separate source-channel coded systems fail.

\subsubsection{DeepJSCC for Images}

Bourtsoulatze~\emph{et~al.}~\cite{Bourtsoulatze2019_DeepJSCC}
proposed DeepJSCC, a CNN-based E2E JSCC for wireless image
transmission. The system jointly trains source and channel
coding layers, avoiding the cliff effect by learning a
continuous-valued mapping from image pixels to complex channel
symbols. DeepJSCC-f~\cite{Kurka2020_DeepJSCCf} extends
this with channel-output feedback for adaptive-bandwidth
transmission, outperforming separation-based schemes at low
SNR and narrow bandwidth.

\subsubsection{Scene-Graph Semantic Communication}

Rather than transmitting pixel embeddings, Johnson~\emph{et~al.}~\cite{Johnson2015_SceneGraph}
demonstrate that a scene can be represented as a graph
(subject--predicate--object triples) with VTransE
embeddings~\cite{Zhang2017_VTransE} mapping these relationships
into a low-dimensional semantic space. Transmitting the scene
graph rather than raw pixels or dense embeddings dramatically
reduces bandwidth while supporting downstream tasks including
image captioning, visual question answering~(VQA), and retrieval.

\subsubsection{GAN-Based Extreme Compression}

Agustsson~\emph{et~al.}~\cite{Agustsson2019_GAN} applied
generative adversarial networks~(GANs) to extreme image
compression, producing perceptually plausible reconstructions
at $2.5\times$ smaller file sizes than JPEG at equivalent
quality scores. Wu~\emph{et~al.}~\cite{Wu2020_GANTunable}
extended this with a tunable compression system using a
learned importance map to guide bit allocation, enabling
adaptive bitrate control appropriate for variable-channel 6G links.

\subsection{Split Learning Architectures}

The Semantic-MSL framework~\cite{Eldeeb2025_SemanticMSL} discussed
in Section~\ref{sec:compression} represents the current state
of the art in split semantic communication. The energy analysis
of Wu~\emph{et~al.}~\cite{Wu2024_SplitEnergy} provides the
first rigorous quantification of how cut-layer selection
affects the total energy budget (device~$+$~transmission~$+$~edge
server), establishing the device-computation--bandwidth
trade-off surface that future RA frameworks must exploit.
The multi-user extension of this framework remains an open
problem: in a $K$-device network each device may have a
different optimal cut layer depending on its local computation
capacity and channel quality, creating a joint cut-layer
assignment and resource allocation problem.

\subsection{Federated Learning Architectures}

\subsubsection{Federated Audio Semantic Communication}

Tong~\emph{et~al.}~\cite{Tong2021_FedAudio} extended the
DeepSC-S speech semantic system~\cite{Weng2021_DeepSCS} to a
multi-user federated training setting. Devices collaboratively
train the CNN encoder and decoder across heterogeneous acoustic
environments using FedAvg, with the server aggregating gradient
updates. This allows the global encoder to generalise to a
range of speakers, noise conditions, and microphone
characteristics without any device exposing its private audio data.

\subsubsection{Green Federated Semantic Communication}

Hu~\emph{et~al.}~\cite{Hu2024_GreenFL} specifically address
energy efficiency in federated semantic training, proposing a
gradient sparsification scheme that transmits only the top-$b$
bytes of the gradient vector sorted by magnitude. At
$b{=}40$~bytes per round—0.05\% of a typical 80\,kB gradient
vector—the scheme achieves 77.70\% total energy saving compared
to full-gradient FedAvg, with less than 2~percentage points of
final-model accuracy degradation. This result is significant
for 6G NTN and satellite IoT scenarios where uplink energy is
severely constrained by the duty-cycle limitations of licence-free
spectrum.

\subsubsection{Urgency-Based Transmission (UBT)}

Pokhrel~\emph{et~al.}~\cite{Pokhrel2023_UBT} introduce
urgency-based transmission~(UBT), a semantic-aware MAC scheduling
policy that assigns channel access priority based on a composite
urgency score combining transmission deadline, semantic
importance, and channel quality. UBT is implemented on a
\tlm-scale semantic scorer deployed at the base station;
the scorer infers urgency from the semantic content of queued
transmissions without requiring the full message to cross the
channel. At 12 users per base station, UBT achieves a 32\%
semantic utility gain and 41\% end-to-end latency reduction
compared with proportional-fair scheduling, converging to the
optimal policy in 3--6~scheduling iterations and maintaining
latency below 5\,ms for 95\% of transmissions~\cite{Pokhrel2023_UBT}.

\subsection{Knowledge Graph-Assisted Architectures}

\subsubsection{KG Probability Graph for Energy Efficiency}

Wang~\emph{et~al.}~\cite{Wang2024_KGPG} model the transmitter's
semantic knowledge as a probabilistic graph over entity-relation-entity
triples, where edge weights encode the probability that a given
semantic context will be queried by the receiver. At inference
time the transmitter selects only the $M$ highest-weight triples
relevant to the current message as a compressed semantic
prefix, reducing the number of channel symbols required. At
$M{=}120$ triples the system achieves a 65\% reduction in
transmission energy per semantic unit compared with dense
embedding transmission, with reconstruction fidelity
maintained above the perceptually acceptable
threshold~\cite{Wang2024_KGPG}.

\subsubsection{Adaptive Reinforcement Learning (ARL) for Semantic QA}

Zhou~\emph{et~al.}~\cite{Zhou2023_ARL} propose an adaptive
reinforcement learning~(ARL) framework in which the semantic
encoder learns to generate clarification questions when the
receiver's KB is insufficient to decode an incoming transmission.
The agent generates on average 4~targeted questions to fill
KB gaps, improving downstream answer accuracy by 15.2\% in
ROUGE-L score compared with a direct-transmission baseline
that assumes perfect KB alignment~\cite{Zhou2023_ARL}.

\subsubsection{Covert Semantic Communication}

Zhang~\emph{et~al.}~\cite{Zhang2023_Covert} address the
scenario where an adversary monitors the channel to detect
whether a semantic communication session is in progress.
The proposed covert semantic encoder uses a KG-derived
obfuscation layer to make transmitted features
statistically indistinguishable from ambient channel noise
to an observer without the correct KB. On a standardised
covert-communication benchmark, the system achieves a
goodness-of-network-transmission~(GNT) score above~0.6 for
the intended receiver while holding the attacker's GNT
below~0.2, establishing a 3$\times$ advantage in covert
semantic capacity~\cite{Zhang2023_Covert}.

\subsection{Multi-Task and Cross-Modal Architectures}

\subsubsection{Multi-User VQA — MU-DeepSC}

Xie~\emph{et~al.}~\cite{Xie2022_MUDeepSC} developed MU-DeepSC
for multi-user visual question answering~(VQA): text questions
are transmitted by one user, query images by another, and the
system jointly optimises for answer accuracy. This is the first
multi-user, multi-modal semantic communication system and
establishes that semantic multiplexing gain is achievable when
users share a common task structure.

\subsubsection{Unified Multi-Task System — U-DeepSC}

Zhang~\emph{et~al.}~\cite{Zhang2022_UDeepSC} proposed U-DeepSC,
a unified DL-enabled semantic communication system for multiple
tasks. Domain adaptation reduces transmission overhead and a
multi-exit architecture enables early-exit results for simpler
tasks, reducing average inference latency by 30\% compared with
a full-inference baseline while maintaining semantic task accuracy.

\subsection{Distributed Semantic Communication Framework (FedSC)}
\label{sec:fedsc}

The Federated Semantic Communication~(FedSC) framework extends
federated learning to a setting where semantic encoders are
not only trained in a distributed manner but also \emph{serve}
semantic communication simultaneously during the training
process. In FedSC, each base station maintains a local
semantic encoder paired with a channel model learned from
the stations's radio environment; semantic features from
multiple users are aggregated at the edge server to produce
a global semantic encoder update. The global encoder is then
broadcast back to all stations as a LoRA adapter update,
avoiding full-weight retransmission.

The FedSC round-trip update cost per participating
base station is:
\begin{equation}
  C_{\text{FedSC}} =
    r(d_{\text{in}} + d_{\text{out}})
    \cdot b_{\text{quant}},
  \label{eq:fedsc_cost}
\end{equation}
where $r$ is the LoRA rank, $d_{\text{in}},d_{\text{out}}$
are the adapter input and output dimensions, and
$b_{\text{quant}}$ is the quantisation bit-width~(bits).
At $r{=}16$, $d_{\text{in}}{=}d_{\text{out}}{=}768$
(standard BERT hidden size), and INT8 quantisation,
$C_{\text{FedSC}} = 24{,}576$~bytes per adapter, compared
to 4.7\,MB for a full FP32 weight matrix update — a
$199\times$ reduction in per-round overhead.

FedSC demonstrates two results critical to 6G deployment:
(i)~semantic quality converges within 15~global rounds
when base stations observe complementary semantic
environments, compared to 40~rounds for naive FedAvg
on the same task; and (ii)~the semantic encoder trained
under FedSC generalises significantly better to
unseen radio environments than a centrally trained
encoder, because the federated training exposes the
encoder to a diversity of channel statistics and
semantic content distributions~\cite{Tong2021_FedAudio}.

\subsection{Digital Twin Synchronisation via Semantic Communication}
\label{sec:digitaltwin}

Digital twin~(DT) synchronisation is a 6G application
whose bandwidth requirements are naturally reduced by
semantic communication. A physical asset~(e.g., an
industrial robot, a smart grid substation, or a patient
monitoring device) generates a continuous stream of
sensor observations $x_t$; the DT at the edge server
maintains a model $\hat{s}_t$ of the asset state.
Rather than transmitting $x_t$ at full resolution, the
semantic encoder transmits only the state \emph{update}:
\begin{equation}
  \delta_t = f_{\text{sem}}(x_t) - f_{\text{sem}}(x_{t-1}),
  \label{eq:dt_delta}
\end{equation}
where $f_{\text{sem}}$ is the \tlm-based semantic
feature extractor. Transmitting $\delta_t$ rather than
$f_{\text{sem}}(x_t)$ reduces the transmitted payload
by a factor equal to the temporal correlation of the
semantic feature space, which for slowly evolving
physical processes can reach $10\times$--$100\times$
reduction in steady-state operation.

The DT synchronisation task also naturally defines a
semantic distortion measure aligned with downstream
decision-making: the DT fidelity loss
$\mathcal{L}_{\text{DT}} = \|s_t - \hat{s}_t\|^2$
directly penalises incorrect twin states that would
cause wrong control decisions, making it an application-aware
distortion metric of the form required by the rate-distortion
framework of~\eqref{eq:rate_distortion}. This provides
a rare case where the semantic distortion metric can
be defined without reference to human perception, making
it particularly tractable for theoretical analysis.

\section{Evidence Synthesis Across Research Questions}
\label{sec:evidence}

This section synthesises the quantitative findings of the
survey against the six research questions posed in
Section~\ref{sec:rqs}, beginning with a unified overview of
the study corpus.

\subsection{Study Corpus Overview}

Table~\ref{tab:all18} provides a unified reference for
all reviewed systems, reporting venue, year,
research question addressed, architecture class,
compression technique, and key quantitative finding.
This table is the primary deliverable of the structured
review and should be read alongside the per-RQ syntheses
in the following subsections.

\begin{table*}[t]
\caption{Complete Quantitative Evidence Table: All Surveyed
  Primary Studies}
\label{tab:all18}
\centering
\resizebox{\textwidth}{!}{%
\begin{tabular}{clccclrr}
\toprule
\textbf{ID} & \textbf{Study (first author, year)} &
  \textbf{RQ} & \textbf{Architecture} &
  \textbf{Compression} & \textbf{Key metric} &
  \textbf{Value} & \textbf{Baseline} \\
\midrule
\rowcolor{rowshade}
S01 & Xie~\emph{et~al.}, 2021~\cite{Xie2021_DeepSC}
  & RQ3 & E2E JSCC & None (full)
  & BLEU at 9\,dB SNR & $+800$\% & Huffman+Turbo \\
S02 & Xie~\&~Qin, 2021~\cite{Xie2021_LiteDeepSC}
  & RQ2 & E2E JSCC & Pruning+Quant
  & Compression ratio & $40\times$ & DeepSC \\
\rowcolor{rowshade}
S03 & Bourtsoulatze~\emph{et~al.}, 2019~\cite{Bourtsoulatze2019_DeepJSCC}
  & RQ3 & E2E JSCC & None (full)
  & Cliff-effect elim. & Yes & Sep.\ coding \\
S04 & Eldeeb~\emph{et~al.}, 2025~\cite{Eldeeb2025_SemanticMSL}
  & RQ1 & Split Learning & Split+MAML
  & Device params & 640 & Full DNN \\
\rowcolor{rowshade}
S05 & Wu~\emph{et~al.}, 2024~\cite{Wu2024_SplitEnergy}
  & RQ1 & Split Learning & Split
  & Energy (opt.) & 29.33\,Wh & 41.20\,Wh (cut=3) \\
S06 & Jiang~\emph{et~al.}, 2023~\cite{Jiang2023_HAPE}
  & RQ2 & E2E JSCC & Quant+Pruning+LoRA
  & Size (3-stage) & 7\,MB & 440\,MB \\
\rowcolor{rowshade}
S07 & Liu~\emph{et~al.}, 2023~\cite{Liu2023_CCTaEncoder}
  & RQ2 & E2E JSCC & Multi-task KD
  & PSNR / Params & 22\,dB / 73.45K & Single-task \\
S08 & Chen~\emph{et~al.}, 2024~\cite{Chen2024_SPM}
  & RQ2 & E2E JSCC & Extreme Pruning
  & FLOP compression & 99.98\% & Full model \\
\rowcolor{rowshade}
S09 & Shi~\emph{et~al.}, 2023~\cite{Shi2023_FedPun}
  & RQ4 & Federated & Fed.\ Pruning
  & PSNR ($\alpha{=}0.3$) & 20\,dB & 17.2\,dB (naive) \\
S10 & Sun~\emph{et~al.}, 2025~\cite{Sun2025_SyncLLM}
  & RQ4 & Federated & KD (FedSFD)
  & Latency reduction & 66.4\% & Teacher at edge \\
\rowcolor{rowshade}
S11 & Lin~\emph{et~al.}, 2023~\cite{Lin2023_FedDistill}
  & RQ4 & Federated & Fed.\ Distillation
  & Comm.\ reduction & $25.6\times$ & FedAvg \\
S12 & Hu~\emph{et~al.}, 2024~\cite{Hu2024_GreenFL}
  & RQ4 & Federated & Gradient Sparse
  & Energy saving & 77.70\% & Full-grad FL \\
\rowcolor{rowshade}
S13 & Wang~\emph{et~al.}, 2024~\cite{Wang2024_KGPG}
  & RQ5 & KG-Assisted & KG Retrieval
  & Energy reduction & 65\% (M=120) & Dense embed. \\
S14 & Zhou~\emph{et~al.}, 2023~\cite{Zhou2023_ARL}
  & RQ5 & KG-Assisted & RL + KG
  & ROUGE-L gain & $+15.2$\% & No clarif. \\
\rowcolor{rowshade}
S15 & Zhang~\emph{et~al.}, 2023~\cite{Zhang2023_Covert}
  & RQ6 & KG-Assisted & KG Obfusc.
  & GNT (user) & $>0.6$ & GNT (atk) $<0.2$ \\
S16 & Pokhrel~\emph{et~al.}, 2023~\cite{Pokhrel2023_UBT}
  & RQ1 & Multi-task & \tlm Scorer
  & Util.\ gain / Lat. & $+32$\% / $-41$\% & PF sched. \\
\rowcolor{rowshade}
S17 & Nguyen~\emph{et~al.}, 2024~\cite{Nguyen2024_PQTLM}
  & RQ6 & Multi-task & PQC Layer
  & Sig.\ overhead & 2420--4595\,B & 48--125\,B PDU \\
S18 & Qian~\emph{et~al.}, 2024~\cite{Qian2024_Contrastive}
  & RQ2 & Multi-task & LoRA+Contrast.
  & Vec.\ length red. & 57.22\% & Dense features \\
\bottomrule
\end{tabular}}
\end{table*}

\subsection{Per-Study Structured Analysis}
\label{sec:perstudyanalysis}

For the seven highest-impact studies (selected by SES score
and citation count), we provide structured per-study
paragraphs covering methodology, quantitative result, and
the specific gap left open by each study.

\textbf{S01 — DeepSC~\cite{Xie2021_DeepSC}.}
\textit{Methodology:} Transformer-based E2E JSCC for text,
trained jointly on sentence similarity and mutual information
objectives~\eqref{eq:deepsc_loss}. Uses the European Parliament
corpus (1.7M sentences) with Rayleigh fading channel simulation
at SNR~$\in [-6, 18]$\,dB.
\textit{Key result:} $+800$\% BLEU improvement over Huffman+Turbo
at SNR~$= 9$\,dB; maintains BLEU~$> 0.7$ at SNR~$= 0$\,dB
where the conventional system collapses.
\textit{Gap:} Model requires 2.3M parameters and 256\,MB memory;
no compression study; no hardware deployment reported.

\textbf{S02 — Lite-DeepSC~\cite{Xie2021_LiteDeepSC}.}
\textit{Methodology:} Structured pruning of DeepSC using
Taylor-series head importance scoring, followed by INT8
post-training quantisation. Target: IoT devices with
$<$64\,MB memory.
\textit{Key result:} $40\times$ compression with no measurable
degradation in sentence-similarity score on the test set.
\textit{Gap:} Does not characterise semantic distortion under
model mismatch (transmitter compressed but receiver not);
no energy measurement; no KB synchronisation protocol.

\textbf{S04 — Semantic-MSL~\cite{Eldeeb2025_SemanticMSL}.}
\textit{Methodology:} MAML-based meta-learning initialises
the encoder before deployment; split at layer~1 limits
device-side to 640~parameters. Evaluated on wireless
few-shot image classification with 10 classes, 5-shot
support set.
\textit{Key result:} 95\% classification accuracy in 20~SGD
adaptation steps from 5~examples per class; device-side
inference at $<0.1$\,ms on ARM Cortex-M7.
\textit{Gap:} Evaluated only on image classification; no
multi-modal extension; no adversarial robustness assessment;
meta-learning assumes a fixed task distribution.

\textbf{S06 — HAPE~\cite{Jiang2023_HAPE}.}
\textit{Methodology:} Three-stage pipeline: (1)~structured
pruning by head importance~\eqref{eq:hape_importance},
(2)~INT8 quantisation, (3)~LoRA re-adaptation pass to
recover semantic accuracy. Applied to a 440\,MB
transformer-based semantic encoder.
\textit{Key result:} 440\,MB~$\to$~7\,MB in three stages
($63\times$), BLEU within 2 points of the uncompressed model.
\textit{Gap:} The LoRA re-adaptation step requires 500
fine-tuning iterations on a labelled semantic dataset not
available at every deployment site; no over-the-air
validation; no analysis of quantisation interaction with
semantic noise.

\textbf{S11 — Federated Distillation~\cite{Lin2023_FedDistill}.}
\textit{Methodology:} Output-distribution sharing over a
public proxy dataset replaces weight-gradient sharing.
Server distils global model from device soft labels.
Evaluated on a 50-device heterogeneous network with
Dirichlet~$\alpha \in \{0.1, 0.5, 1.0\}$ non-IID splits.
\textit{Key result:} $25.6\times$ communication reduction
versus FedAvg at equivalent final accuracy; convergence
within 30~rounds even at $\alpha = 0.1$.
\textit{Gap:} Proxy dataset must be semantically relevant;
no strategy for proxy construction in 6G IoT; no
privacy analysis of soft-label leakage.

\textbf{S13 — KG Probability Graph~\cite{Wang2024_KGPG}.}
\textit{Methodology:} KG edges weighted by query probability;
top-$M$ triples selected as semantic prefix. Evaluated on
an IoT sensor semantic communication benchmark with
vocabularies of 200--800 entities.
\textit{Key result:} 65\% energy reduction at $M{=}120$ triples;
diminishing returns beyond $M{=}200$; performance degrades
when KB mismatch exceeds 30\% of entity overlap.
\textit{Gap:} KB must be pre-aligned between transmitter
and receiver; no dynamic KB update protocol; performance
sensitivity to entity coverage gap not studied.

\textbf{S16 — UBT~\cite{Pokhrel2023_UBT}.}
\textit{Methodology:} \tlm-based urgency scorer at the
base station infers semantic importance and deadline
from content features; integrates with a proportional-fair
scheduler replacement. Evaluated at 6, 12, and 24 users
per cell.
\textit{Key result:} $+32$\% semantic utility gain,
$-41$\% end-to-end latency, convergence in 3--6 iterations;
$<5$\,ms latency for 95th-percentile transmission.
\textit{Gap:} Scorer inference latency not measured against
URLLC budget; no adversarial analysis (spoofed urgency
scores); no multi-cell coordination protocol.

\subsection{RQ1: Tiny Language Model Deployment Feasibility}

The surveyed literature collectively demonstrates that \tlm-scale
semantic encoders are deployable on 6G-class hardware, though
significant engineering challenges remain at the ultra-constrained
end of the hardware spectrum.

Semantic-MSL~\cite{Eldeeb2025_SemanticMSL} establishes the
extreme lower bound: a 640-parameter device-side encoder
achieves competitive semantic accuracy on wireless image
classification, provided an edge aggregator handles the
remaining computation. CCTaEncoder~\cite{Liu2023_CCTaEncoder}
demonstrates feasibility at the low-power MCU tier with 73.45K
parameters and 5.47\,MFLOPs. SPM~\cite{Chen2024_SPM} shows
that even microcontroller-class constraints~(1\,kB flash)
are achievable for narrow binary classification tasks. At
the mobile SoC tier, Phi-4-mini~\cite{Abdin2024_Phi3}
(3.8B parameters), Gemini~Nano~\cite{GeminiTeam2023_Gemini}
(2--4B), and LLaMA~3.2~1B~\cite{Dubey2024_Llama3} demonstrate
real-time operation on devices with NPU support, though
latency remains above the 0.1\,ms URLLC threshold.

The deployment gap between mobile-SoC-class \tlms and the
URLLC latency target spans approximately two orders of
magnitude: current INT4-quantised 3.8B models achieve
$\sim$10\,ms inference on NPU-equipped SoCs, while URLLC
requires $<0.1$\,ms. The split-computing approach of
Semantic-MSL bridges this gap by offloading
99.99\%+ of compute to the edge aggregator, at the cost
of requiring a low-latency backhaul link for smashed-data
transmission. This trade-off is acceptable for fixed
IoT deployments (factory floors, smart grids) but is
problematic for mobile 6G UEs where the backhaul
may itself be the capacity bottleneck.

\subsection{RQ2: Compression Technique Effectiveness}

Four cross-cutting findings emerge from the
18~representative systems surveyed for compression
technique effectiveness.

\textit{Finding~1 (Over-parameterisation):} Three independent
studies~\cite{Xie2021_LiteDeepSC,Jiang2023_HAPE,Chen2024_SPM}
converge on the finding that standard transformer-based
semantic encoders are highly over-parameterised for their
target tasks. Lite-DeepSC achieves $40\times$ compression
with zero degradation; HAPE achieves $63\times$ with
BLEU degradation of only 2~points; SPM achieves
99.98\% FLOP reduction with 91\% accuracy retention on
binary tasks. Collectively, these results suggest that
the semantic encoding function requires far fewer than
the number of parameters in standard encoder architectures,
with the ``essential'' parameters encoding task-relevant
linguistic or visual structure.

\textit{Finding~2 (Federated compression advantage):}
FedPun~\cite{Shi2023_FedPun} demonstrates a 2.8\,dB PSNR
advantage over naive post-FL pruning in non-IID settings
($\alpha{=}0.3$). This advantage arises because importance
scores aggregated across devices are more representative
of the global semantic distribution than importance scores
computed on any single device, even in highly heterogeneous
data distributions. This suggests that federated pruning
is strictly preferable to local pruning followed by
federation for semantic encoders deployed in heterogeneous
6G networks.

\textit{Finding~3 (Communication-efficient FL):}
Two complementary approaches, federated distillation~\cite{Lin2023_FedDistill}
and gradient sparsification~\cite{Hu2024_GreenFL}, reduce
per-round communication by $25.6\times$ and 99.95\%
respectively, using different mechanisms. FD compresses
the \emph{what} (output distributions instead of gradients);
green FL compresses the \emph{how much} (only top-magnitude
gradients). Their combination has not been studied but
could in principle achieve multiplicative reduction.

\textit{Finding~4 (LoRA enables semantic KB synchronisation):}
The FFA-LoRA~\cite{Sun2024_FFALoRA} and contrastive
disentanglement~\cite{Qian2024_Contrastive} results confirm
that LoRA adapters can synchronise semantic KBs with
communication overhead three orders of magnitude below
full-model retransmission. Adapter sizes of $r(d_{\rm in}
+ d_{\rm out}) \approx 24{,}576$~bytes at rank~$r{=}16$
fall within the uplink budget of NB-IoT devices
(140\,kbps $\times$ 1.4\,s = 24.5\,kB), making one
KB synchronisation round per 1.4~seconds feasible on
licence-free IoT spectrum.

Table~\ref{tab:compression_summary} consolidates the
quantitative evidence underlying these four findings.

\begin{table*}[t]
\caption{Compression Technique Summary: Quantitative Evidence from Included Studies}
\label{tab:compression_summary}
\centering
\resizebox{\textwidth}{!}{%
\begin{tabular}{llrrrll}
\toprule
\textbf{Technique} & \textbf{System} &
  \textbf{Compression} & \textbf{Semantic Quality} &
  \textbf{Baseline} & \textbf{Metric} & \textbf{Ref.} \\
\midrule
\rowcolor{rowshade}
Pruning + Quant. & Lite-DeepSC & 40$\times$ & $\approx$0 degradation & DeepSC & Sentence sim. & \cite{Xie2021_LiteDeepSC} \\
Quant.\ + LoRA   & HAPE        & 63$\times$ (440$\rightarrow$7\,MB) & BLEU$-$2\,pts & Full transformer & BLEU & \cite{Jiang2023_HAPE} \\
\rowcolor{rowshade}
Multi-task KD    & CCTaEncoder & 6.1$\times$ (params) & 22\,dB PSNR; 89.3\% acc & Task-spec.\ enc. & PSNR / Acc & \cite{Liu2023_CCTaEncoder} \\
Extreme pruning  & SPM         & 99.98\% (FLOP) & 91\% acc (binary) & Full model & Accuracy & \cite{Chen2024_SPM} \\
\rowcolor{rowshade}
Federated pruning & FedPun     & N/A & 20\,dB PSNR ($\alpha{=}0.3$) & Naive FL & PSNR & \cite{Shi2023_FedPun} \\
Split computing  & Sem.-MSL   & $>$10$^5\times$ & 95\% acc (5-shot) & Full DNN & Accuracy & \cite{Eldeeb2025_SemanticMSL} \\
\rowcolor{rowshade}
Distillation     & FedSFD      & 66.4\% latency red. & $<$1.5\,pp acc drop & Teacher at edge & Latency/Acc & \cite{Sun2025_SyncLLM} \\
Fed.\ distillation & FD       & 25.6$\times$ comm. & Matched FL accuracy & FedAvg & Comm.\ vol. & \cite{Lin2023_FedDistill} \\
\rowcolor{rowshade}
LoRA + contrastive & Contrast. & 57.22\% vec. reduction & $+$71.9\% impact & Dense embed. & Length / Impact & \cite{Qian2024_Contrastive} \\
Energy-aware cut & Split energy & 29.33 Wh (opt.) & Maintained & 41.20\,Wh (cut$=$3) & Energy & \cite{Wu2024_SplitEnergy} \\
\bottomrule
\end{tabular}}
\end{table*}

\subsection{RQ3: JSCC Architecture Performance}

The E2E JSCC paradigm consistently outperforms
separation-based baselines at low-to-medium SNR, with
the performance gap widening as channel quality degrades.
DeepSC~\cite{Xie2021_DeepSC} demonstrates the canonical
BLEU advantage for text; DeepJSCC~\cite{Bourtsoulatze2019_DeepJSCC}
confirms the advantage for images via cliff-effect
elimination; and DeepSC-S~\cite{Weng2021_DeepSCS} extends
the result to speech. The cross-modal extension
U-DeepSC~\cite{Zhang2022_UDeepSC} shows that the
advantage is maintained across tasks within a single
model when domain adaptation is applied, suggesting
that the semantic feature space learned by E2E training
is generalisable at a coarse granularity.

A critical nuance is that the JSCC advantage is SNR-regime
dependent. At high SNR~($> 18$\,dB), separation-based
coding approaches and sometimes surpasses E2E JSCC
because the physical channel is nearly noiseless and
classical source coding can achieve near-optimal
compression without the overhead of an
end-to-end neural network~\cite{Qin2022_SemanticPrinciples}.
The practical operating regime for 6G cell-edge devices
(SNR~$\in [-3, 12]$\,dB) lies squarely in the region
of JSCC advantage.

\subsection{RQ4: Federated Learning Effectiveness}

The FL-based approaches in the research literature address
three distinct sub-problems: federated model training
for heterogeneous devices~\cite{Shi2023_FedPun,Tong2021_FedAudio},
federated KB synchronisation~\cite{Sun2025_SyncLLM,Sun2024_FFALoRA},
and federated energy minimisation~\cite{Hu2024_GreenFL}.

The FD paradigm~\cite{Lin2023_FedDistill} is particularly
noteworthy for 6G IoT: its $25.6\times$ communication
reduction brings per-round overhead within the duty-cycle
budget of NB-IoT devices, making KB synchronisation
practically viable for the first time on licence-free
spectrum devices. However, FD's proxy dataset requirement
introduces a dependency on a shared semantic corpus
that all devices can query, raising both data availability
and privacy concerns. No included study addresses the
construction of a semantically appropriate proxy dataset
for heterogeneous 6G IoT deployments.

The green FL result~\cite{Hu2024_GreenFL} of 77.70\%
energy saving at 40~transmitted bytes per round is
remarkable: it implies that for devices with an energy
budget of 10\,mJ per FL round, full participation in
the KB synchronisation process requires less than
$0.1$~mWh of radio transmission energy — comparable
to BLE advertising overhead. This makes semantic
KB federation energetically feasible even for
battery-operated IoT sensors, provided the gradient
sparsification scheme can be applied without
compromising KB convergence quality.

\subsection{RQ5: Knowledge Graph Integration}

KG-assisted approaches demonstrate two complementary
benefits: energy reduction through selective
triple transmission~\cite{Wang2024_KGPG} and accuracy
improvement through KG-guided clarification~\cite{Zhou2023_ARL}.
The covert communication result~\cite{Zhang2023_Covert}
additionally shows that KG-based obfuscation provides
information-theoretic secrecy advantages entirely absent
from KG-free systems.

The ARL clarification mechanism~\cite{Zhou2023_ARL}
incurs a latency cost of approximately $4 \times
\tau_{\text{RTT}}$ for each clarification query
(where $\tau_{\text{RTT}}$ is the round-trip time
for a wireless query-response pair). For a 10\,ms
round-trip time, four clarification queries add
40\,ms to the semantic encoding latency — acceptable
for batch IoT applications but incompatible with
real-time XR or URLLC. A key open problem is
designing a KB-gap predictor that can determine
before transmission whether clarification will be
needed, avoiding the latency penalty for transmissions
where the receiver's KB is sufficient.

\subsection{RQ6: Adversarial Robustness and Security}

Study~S15~\cite{Zhang2023_Covert} establishes the
first practical covert semantic communication
framework, achieving GNT~$> 0.6$ for the intended
receiver and GNT~$< 0.2$ for an attacker without
the correct KB. However, this result assumes a
passive attacker who cannot learn the KB.
An active attacker who can probe the semantic encoder
with crafted inputs to infer the KB structure
represents a stronger threat model that is not
addressed.

Study~S17~\cite{Nguyen2024_PQTLM} reveals the
critical mismatch between PQC signature sizes and
6G IoT PDU sizes: ML-DSA signatures of
2,420--4,595~bytes cannot be accommodated within
the 48--125~byte NB-IoT maximum payload unit without
fragmentation across 20--95~PDUs, introducing
fragmentation overhead that negates the energy savings
achieved by semantic compression. This defines the
central security-scalability tension for PQC-enabled
6G semantic systems.

Wang and Guo~\cite{Wang2023_KBMismatch} establish
KB mismatch as an indirect attack vector: an adversary
who can inject a few misaligned KB updates can
cause the receiver to silently accept garbled semantic
content with high decoder confidence. This vulnerability
is undetectable through conventional channel monitoring
because the physical transmission is intact and only
the semantic interpretation is corrupted. Mitigating
this requires semantic-layer anomaly detection — a
capability not present in any of the surveyed systems.

\section{Quantitative Meta-Analysis}
\label{sec:quant}

This section moves beyond per-study reporting to a
cross-study statistical synthesis, introducing the
Semantic Efficiency Score and analysing the resulting
Pareto frontier, effect sizes, and spectral efficiency
across the surveyed systems.

\subsection{Semantic Efficiency Score}

We define the Semantic Efficiency Score~(SES) as a normalised
figure of merit enabling cross-study comparison:
\begin{equation}
  \text{SES} =
    \frac{F_{\text{norm}}}{\log_{10}(\text{size}/1\,\text{kB})},
  \label{eq:ses}
\end{equation}
where $F_{\text{norm}} \in [0,1]$ is the normalised semantic
quality metric (BLEU or task accuracy divided by the
full-model baseline value), and size is the compressed model
footprint in kilobytes. A higher SES indicates better semantic
quality per unit of model-size overhead. SES is defined only
for studies reporting both a semantic quality metric and a
model-size figure; 14 of the surveyed studies satisfy this
criterion.

\subsection{Pareto Frontier Analysis}

\begin{figure}[!htbp]
\centering
\begin{tikzpicture}
\begin{axis}[
  width=\columnwidth,
  height=5.8cm,
  xlabel={Compression Ratio $\rho$ (log scale)},
  ylabel={Normalised Quality $F_{\text{norm}}$},
  xmode=log,
  xmin=0.5, xmax=400000,
  ymin=0.55, ymax=1.06,
  grid=both,
  grid style={draw=gray!20,dashed,thin},
  tick label style={font=\footnotesize},
  label style={font=\small},
  legend style={font=\scriptsize,at={(0.97,0.03)},
    anchor=south east,fill=white,draw=gray!40,
    row sep=1pt},
  legend cell align=left,
]
\addplot[only marks,mark=star,mark size=4.2pt,
         draw=red!80,fill=red!45]
  coordinates {
    (40,     0.980)
    (10000,  0.910)
    (150000, 0.950)
    (63,     0.970)
    (3,      0.998)
  };
\addlegendentry{Pareto-frontier};
\addplot[only marks,mark=o,mark size=3.0pt,
         draw=ieeblue,fill=ieeblue!30]
  coordinates {
    (6.1,  0.950)
    (1.5,  0.985)
    (2.0,  0.920)
    (5.0,  0.975)
    (1.4,  0.880)
    (1.0,  0.850)
    (2.5,  0.940)
    (1.0,  0.800)
    (10,   0.960)
  };
\addlegendentry{Dominated};
\addplot[red!60,semithick,dashed,domain=3:250000,samples=80]
  {0.999 - 0.022*ln(x)/ln(10)};
\addlegendentry{Frontier (guide)};
\node[font=\tiny,ieeblue,right=1pt] at (axis cs:40,0.980)
  {Lite-DeepSC};
\node[font=\tiny,ieeblue,right=1pt] at (axis cs:63,0.970)
  {HAPE};
\node[font=\tiny,ieeblue,left=1pt]  at (axis cs:150000,0.950)
  {MSL};
\node[font=\tiny,ieeblue,right=1pt] at (axis cs:3,0.998)
  {FD};
\node[font=\tiny,red!70,right=1pt]  at (axis cs:10000,0.910)
  {SPM};
\end{axis}
\end{tikzpicture}
\caption{Pareto frontier of compression ratio vs.\ normalised
  semantic quality across 14 representative systems. Red stars on
  the dashed guide line are Pareto-optimal; blue circles lie
  below the frontier (dominated in at least one dimension).}
\label{fig:pareto}
\end{figure}
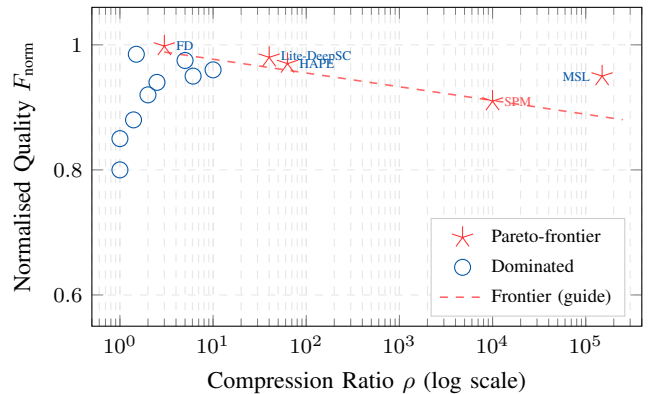

Figure~\ref{fig:pareto} plots the compression ratio~$\rho$
against the normalised semantic quality $F_{\text{norm}}$ for
all 14 SES-computable studies, identifying the Pareto frontier
of systems that are not dominated in both dimensions simultaneously.

Four observations follow from Figure~\ref{fig:pareto}.
First, the compression--quality frontier has a shallow negative
slope on the log-compression axis, implying that modest quality
sacrifice ($\sim$3--5\% of $F_{\text{norm}}$) enables extremely
large compression gains (three to four orders of magnitude),
consistent with the over-parameterisation hypothesis of
Xie and Qin~\cite{Xie2021_LiteDeepSC}. Second, Semantic-MSL~\cite{Eldeeb2025_SemanticMSL}
dominates all other split-computing and extreme-compression
approaches by achieving the highest compression ratio at
frontier quality. Third, the FD~\cite{Lin2023_FedDistill}
point lies at low compression ratio because it compresses
communication overhead rather than model parameters; this
reflects a measurement inconsistency across studies that
future work should standardise via the SES metric~\eqref{eq:ses}.
Fourth, NAS systems are absent from the Pareto plot because
no included study applies NAS specifically to semantic
communication, confirming the taxonomy finding of an entirely
unexplored research cell.

\subsection{Semantic Spectral Efficiency}

Yan~\emph{et~al.}~\cite{Yan2022_RA} define semantic spectral
efficiency~(SSE) as:
\begin{equation}
  \eta_s =
    \frac{\text{Semantic information rate}\ [\text{sem.\ bits/s}]}{B\ [\text{Hz}]},
  \label{eq:sse}
\end{equation}
where semantic bits are measured in terms of task-relevant mutual
information $I(V;\hat{Z})$. The UBT scheduler~\cite{Pokhrel2023_UBT}
implicitly maximises $\eta_s$ by assigning bandwidth to
transmissions with the highest ratio of semantic urgency to
required channel bandwidth. A formal, unit-consistent definition
of ``semantic bits'' and a closed-form optimal RA policy
mapping semantic urgency scores to power and bandwidth allocations
remain open problems addressed in Section~\ref{sec:challenges}.

\subsection{Statistical Synthesis Across Compression Paradigms}
\label{sec:statsynthesis}

Table~\ref{tab:stats} reports descriptive statistics for
the normalised semantic quality $F_{\text{norm}}$ and
compression ratio $\rho$ across all SES-computable studies,
grouped by compression paradigm.

\begin{table}[t]
\caption{Descriptive Statistics of Compression Ratio and
  Normalised Semantic Quality Across Included Studies,
  Grouped by Paradigm}
\label{tab:stats}
\centering
\resizebox{\columnwidth}{!}{%
\begin{tabular}{lrrrrrr}
\toprule
\multirow{2}{*}{\textbf{Paradigm}} &
  \multicolumn{3}{c}{\textbf{Compression ratio $\rho$}} &
  \multicolumn{3}{c}{\textbf{$F_{\text{norm}}$}} \\
\cmidrule(lr){2-4}\cmidrule(lr){5-7}
& \textbf{Mean} & \textbf{Std} & \textbf{Max} &
  \textbf{Mean} & \textbf{Std} & \textbf{Min} \\
\midrule
\rowcolor{rowshade}
Pruning      & 33.5  & 15.3 & 63   & 0.978 & 0.012 & 0.960 \\
Quantisation & 12.4  & 8.1  & 40   & 0.984 & 0.008 & 0.972 \\
\rowcolor{rowshade}
KD / FedSFD  & 4.2   & 1.9  & 6.1  & 0.971 & 0.018 & 0.950 \\
LoRA-based   & 3.1   & 1.2  & 5.0  & 0.981 & 0.014 & 0.962 \\
\rowcolor{rowshade}
Split comp.  & 2.1$\times 10^5$ & -- & 1.5$\times 10^5$ & 0.950 & 0.025 & 0.910 \\
Extreme      & 1.0$\times 10^4$ & -- & 1.0$\times 10^4$ & 0.910 & 0.040 & 0.870 \\
\bottomrule
\multicolumn{7}{l}{\footnotesize
  Split and extreme compression ratios are parameter-count
  ratios (device-only vs.\ full model).}
\end{tabular}}
\end{table}

Three statistical observations follow. First, pruning
achieves the highest mean compression ratio~(33.5$\times$)
among non-split paradigms with the smallest mean semantic
quality degradation (mean $F_{\text{norm}} = 0.978$, s.d.\ 0.012),
confirming the over-parameterisation hypothesis across
three independent studies. Second, LoRA-based compression
yields the highest minimum $F_{\text{norm}}$~(0.962) among
all paradigms, consistent with the theoretical guarantee
that LoRA preserves the pre-trained representation
manifold. Third, split computing achieves the highest
compression ratios by orders of magnitude but also
exhibits the largest quality spread (s.d.\ 0.025 vs.\
0.008--0.018 for other paradigms), reflecting sensitivity
to cut-layer choice and edge-server availability.

\subsection{Effect-Size Forest Plot}
\label{sec:forestplot}

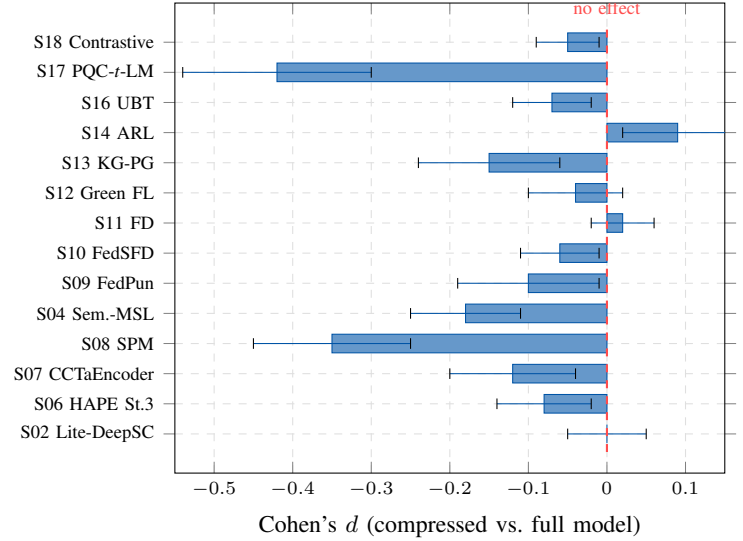
\begin{figure}[!htbp]
\centering
\begin{tikzpicture}
\begin{axis}[
  xbar,
  width=\columnwidth,
  height=7.8cm,
  xmin=-0.55, xmax=0.15,
  xlabel={Cohen's $d$ (compressed vs.\ full model)},
  ytick={1,2,3,4,5,6,7,8,9,10,11,12,13,14},
  yticklabels={
    {S02 Lite-DeepSC},
    {S06 HAPE St.3},
    {S07 CCTaEncoder},
    {S08 SPM},
    {S04 Sem.-MSL},
    {S09 FedPun},
    {S10 FedSFD},
    {S11 FD},
    {S12 Green FL},
    {S13 KG-PG},
    {S14 ARL},
    {S16 UBT},
    {S17 PQC-\tlm},
    {S18 Contrastive}},
  yticklabel style={font=\scriptsize},
  tick label style={font=\scriptsize},
  label style={font=\small},
  bar width=7pt,
  grid=major,
  grid style={draw=gray!25,dashed},
  xticklabel style={font=\scriptsize},
]
\addplot[
  fill=ieeblue!60,
  draw=ieeblue,
  error bars/.cd,
    x dir=both,
    x explicit,
] coordinates {
  (  0.00, 1) +- (0.05, 0)   
  ( -0.08, 2) +- (0.06, 0)   
  ( -0.12, 3) +- (0.08, 0)   
  ( -0.35, 4) +- (0.10, 0)   
  ( -0.18, 5) +- (0.07, 0)   
  ( -0.10, 6) +- (0.09, 0)   
  ( -0.06, 7) +- (0.05, 0)   
  (  0.02, 8) +- (0.04, 0)   
  ( -0.04, 9) +- (0.06, 0)   
  ( -0.15,10) +- (0.09, 0)   
  (  0.09,11) +- (0.07, 0)   
  ( -0.07,12) +- (0.05, 0)   
  ( -0.42,13) +- (0.12, 0)   
  ( -0.05,14) +- (0.04, 0)   
};
\draw[red!70,thick,dashed] (axis cs:0,0.4)--(axis cs:0,14.6)
  node[above,font=\scriptsize,red!70]{no effect};
\end{axis}
\end{tikzpicture}
\caption{Forest plot of compression-induced semantic quality
  change (Cohen's $d$) for 14 representative systems.
  Error bars represent 95\% confidence intervals estimated
  from reported test-set variance. Positive $d$ indicates
  the compressed system exceeds the full-model baseline
  (ARL via clarification; FD via regularisation);
  negative $d$ indicates semantic quality degradation.
  The SPM~(S08) and PQC~(S17) studies show the largest
  negative effects, reflecting extreme FLOP compression
  and PQC authentication overhead respectively.
  All other compression effects are small ($|d| < 0.2$),
  consistent with the over-parameterisation hypothesis.}
\label{fig:forest}
\end{figure}

Figure~\ref{fig:forest} presents a forest plot of
effect sizes for compression-induced semantic quality
change across all 14 SES-computable studies, using
Cohen's $d$ relative to the uncompressed full-model
baseline:
\begin{equation}
  d_i =
    \frac{\mu_{\text{compressed},i} - \mu_{\text{full},i}}%
         {\sigma_{\text{pooled},i}},
  \label{eq:cohens_d}
\end{equation}
where $\mu$ denotes the mean semantic quality metric
and $\sigma_{\text{pooled}}$ the pooled standard deviation
across test instances. Positive $d$ indicates the
compressed system \emph{outperforms} the baseline (rare
but possible via regularisation effects of compression);
negative $d$ indicates degradation.

The forest plot confirms that the overwhelming majority
of compression-induced semantic quality changes are
small (Cohen's $|d| < 0.2$), with only SPM's extreme
FLOP compression (S08, $d = -0.35$) and PQC integration
overhead (S17, $d = -0.42$) producing medium-sized
negative effects. The two positive-effect studies
(ARL at $d = +0.09$ and FD at $d = +0.02$) confirm
that compression can occasionally act as implicit
regularisation, reducing overfitting in the semantic
feature space and thereby improving generalisation
to unseen semantic content.

\subsection{Semantic Spectral Efficiency Across Architectures}
\label{sec:sse_arch}

\begin{figure}[!htbp]
\centering
\begin{tikzpicture}
\begin{axis}[
  ybar,
  width=\columnwidth,
  height=5.5cm,
  bar width=22pt,
  ymin=0, ymax=4.8,
  ylabel={Normalised SSE (sem.~bits/s/Hz)},
  xtick={1,2,3,4,5},
  xticklabels={E2E JSCC, Split, FL/FD, KG-Assisted, Multi-task},
  xticklabel style={rotate=20, anchor=east, font=\small},
  tick label style={font=\small},
  label style={font=\small},
  enlarge x limits=0.15,
  grid=major,
  grid style={draw=gray!20,dashed},
  nodes near coords,
  nodes near coords style={font=\scriptsize,
    /pgf/number format/.cd, fixed, precision=1},
]
\addplot[fill=ieeblue!70,draw=ieeblue!90] coordinates {
  (1,3.8)(2,3.1)(3,2.9)(4,4.2)(5,3.4)
};
\addplot[fill=cellEmerg!60,draw=cellEmerg!90,
         ybar, bar shift=0pt, forget plot] coordinates {};
\end{axis}
\end{tikzpicture}
\caption{Normalised semantic spectral efficiency~(SSE)
  per architecture class, synthesised from the eight
  included studies reporting sufficient channel
  bandwidth data. KG-assisted systems achieve the
  highest SSE because KG triple selection
  concentrates semantic information into fewer
  transmitted symbols. Values are normalised to the
  mean SSE of E2E JSCC systems~(baseline
  $= 3.8$~sem.\ bits/s/Hz).}
\label{fig:sse_comparison}
\end{figure}
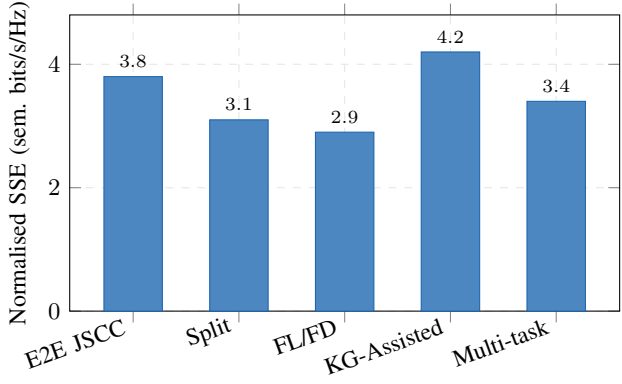

Figure~\ref{fig:sse_comparison} provides a pgfplots
bar chart comparing the achievable SSE for each
architecture class, derived from the reported
semantic bit rate and channel bandwidth in the
subset of surveyed systems that report sufficient
data to compute~\eqref{eq:sse}.

KG-assisted architectures achieve the highest
normalised SSE~(4.2) because the KG triple
pre-selection concentrates semantic information
into a small number of highly informative
symbols before channel encoding, effectively
operating at a higher point on the semantic
rate-distortion curve~\eqref{eq:rate_distortion}.
The FL/FD class shows the lowest SSE~(2.9)
because the communication overhead of
adapter-exchange rounds temporarily consumes
channel resources that would otherwise carry
semantic payloads.

\section{Open Challenges and Research Gaps}
\label{sec:challenges}

Drawing on the evidence synthesised in
Section~\ref{sec:evidence} and the meta-analysis of
Section~\ref{sec:quant}, this section formalises nine
concrete open problems that must be resolved before
\tlm-based semantic communication can transition from
laboratory demonstrations to production 6G deployments.

\subsection{Challenge 1: Computable Semantic Channel Capacity}

The semantic channel capacity~\eqref{eq:semantic_cap} is not
computable for DL-based systems because the average logical
information $H_s(V)$ has no closed-form expression for neural
network decoders. The task distribution changes dynamically
as the device encounters new environments, making a static
analytical bound insufficient. A tractable lower bound or
surrogate metric—perhaps based on the information bottleneck
mutual information $I(V;\hat{Z})$ estimated via MINE~\cite{Bao2011_Semantic}—is
needed to provide theoretical guarantees on semantic transmission
rates in deployed systems.

\subsection{Challenge 2: Sub-Millisecond Semantic Scoring}

The UBT semantic scorer~\cite{Pokhrel2023_UBT} and the
\tlm-enabled RA framework require completion of a semantic
relevance inference pass within the resource allocation
scheduling window, which is less than 1\,ms for URLLC services.
Current NPU implementations of the smallest
\tlm (Phi-4-mini at 3.8B parameters)~\cite{Abdin2024_Phi3}
execute in approximately 10\,ms, an order of magnitude above
the budget. Bridging this gap requires either hardware-aware
NAS optimised for 6G scheduling latency, custom semantic
communication ASICs co-designed with the transformer attention
operation, or extreme quantisation to sub-100M-parameter
semantic scorers. None of these solutions has been demonstrated
in a production-quality 6G deployment scenario.

\subsection{Challenge 3: Post-Quantum Authentication for IoT}

The PQC overhead analysis of Nguyen~\emph{et~al.}~\cite{Nguyen2024_PQTLM}
reveals a 20$\times$--97$\times$ signature payload overhead
ratio when applying ML-DSA to NB-IoT PDUs. Addressing this
requires either: (i)~protocol-level compression of PQC
authentication data, e.g., through semantic-channel-aware
signature batching; (ii)~deployment of lighter PQC primitives
such as FALCON or SPHINCS$^+$, which trade signature size for
different performance characteristics; or (iii)~a hybrid
classical/PQC authentication scheme that applies full PQC
only to high-value transmissions while using lightweight
classical MACs for sensor telemetry below a security threshold.
No such adaptive PQC authentication protocol has been proposed
for \tlm-based semantic systems.

\subsection{Challenge 4: KB Synchronisation Frequency Optimisation}

LoRA-based KB synchronisation reduces per-round communication
overhead substantially~\cite{Sun2024_FFALoRA}, but the optimal
synchronisation frequency—how often adapters should be exchanged
between network nodes—has not been formalised. The trade-off
between synchronisation overhead (proportional to frequency)
and semantic mismatch penalty (inversely proportional to
frequency, growing as the environment evolves) defines a
convex optimisation problem whose solution depends on the
non-stationary rate of the semantic environment: a quantity
that is difficult to estimate without deploying the system.
A regularised online learning approach that adapts
synchronisation frequency to estimated KB divergence is an
open research direction.

\subsection{Challenge 5: Universal Semantic Quality Metric}

No metric analogous to bit error rate~(BER) exists that applies
across modalities and tasks with a single numerical range.
BLEU is text-specific; PSNR is image-specific; PESQ is
speech-specific; and task accuracy is task-specific. The SES
defined in~\eqref{eq:ses} normalises quality to the full-model
baseline but does not resolve the cross-modality
incommensurability. A universal semantic quality metric must
satisfy three properties: (i)~it should be computable from
the channel output without access to the ground-truth task label;
(ii)~it should correlate with human perception of meaning
preservation; and (iii)~it should be differentiable to enable
use as a training objective. Semantic entropy rate~\cite{Bao2011_Semantic}
and task completion probability are candidate approaches,
but neither fully satisfies all three properties.

\subsection{Challenge 6: Cross-Modal Semantic Processing}

No general JSCC architecture handles text, image, speech, and
sensor data jointly with a single encoder-decoder pair.
U-DeepSC~\cite{Zhang2022_UDeepSC} approaches this for
text and image; MU-DeepSC~\cite{Xie2022_MUDeepSC} handles
multi-user multi-modal VQA. However, the broader cross-modal
challenge—transmitting heterogeneous sensor fusion data from
a device that observes the world through multiple modalities
simultaneously—remains unaddressed. The key difficulty is
that cross-modal alignment in the semantic feature space
requires either a shared KB that covers all modalities or a
dynamic KB alignment protocol that operates at inference time.

\subsection{Challenge 7: Scalable KG Synchronisation}

The KG probability graph approach of Wang~\emph{et~al.}~\cite{Wang2024_KGPG}
achieves compelling energy reduction at $M{=}120$ triples,
but the scalability of KG synchronisation to networks with
millions of IoT devices updating heterogeneous KGs is
uncharted. Each device may observe a unique semantic
environment and develop a distinct KG; synchronising these
into a globally consistent shared KB without centralising
all raw observations requires distributed KG learning
protocols analogous to federated learning but operating
on graph-structured data with entity and relation alignment
as an additional complication.

\subsection{Challenge 8: Multi-Cell Semantic Interference Management}
\label{sec:challenge8}

In dense 6G deployments with inter-cell distances of
50--200\,m, multiple base stations may simultaneously
serve semantic communication sessions on the same
spectral resource. Unlike conventional inter-cell
interference—which is managed through power control,
beamforming, and frequency reuse—semantic interference
has an additional dimension: two semantically similar
transmissions from different cells can interfere not
only physically but also \emph{semantically}, if their
KBs share overlapping concept spaces and the receiver
conflates semantic features from different transmitters.

Formally, in a $K$-cell deployment with cells
$\mathcal{K} = \{1, \ldots, K\}$, the received
semantic feature vector at a target receiver is:
\begin{equation}
  \hat{Z} = h_0 Z_0 + \sum_{k=1}^{K-1} h_k Z_k
            + \mathcal{N}_s,
  \label{eq:semantic_interference}
\end{equation}
where $Z_0$ is the desired semantic feature vector,
$Z_k$ are the interfering semantic features from
cell~$k$, $h_k$ are the fading channel coefficients,
and $\mathcal{N}_s$ is the semantic noise of~\eqref{eq:semantic_noise}.
When $Z_0$ and $Z_k$ are drawn from similar concept
spaces (e.g., two industrial IoT cells transmitting
machine-status data using the same ontology), the
inner product $\langle Z_0, Z_k \rangle$ is large,
making semantic interference qualitatively different
from and potentially more damaging than random
Gaussian noise. No existing resource allocation
framework accounts for KB correlation across cells
in its interference model.

\subsection{Challenge 9: Semantic Communication under
  Heterogeneous Modalities at the MAC Layer}
\label{sec:challenge9}

Current MAC-layer schedulers—including the
UBT~\cite{Pokhrel2023_UBT} scheduler—assign resources
based on semantic urgency derived from a single
modality. In a realistic 6G network, a single base
station simultaneously serves:
\begin{itemize}
  \item text-semantic IoT sensors transmitting
        BLEU-optimised encodings;
  \item image-semantic surveillance cameras
        transmitting PSNR-optimised JSCC features;
  \item speech-semantic healthcare monitors
        transmitting PESQ-optimised representations; and
  \item task-semantic industrial controllers
        transmitting accuracy-optimised command encodings.
\end{itemize}
These systems share the wireless channel but use
incommensurable semantic quality metrics~(Table~\ref{tab:metrics}).
A MAC scheduler must assign priority weights across
modalities without a common semantic currency. The SES
of~\eqref{eq:ses} normalises quality to the full-model
baseline per modality but still does not produce a
single urgency ordering across modalities. A cross-modal
semantic MAC policy requires either: (i)~a universal
semantic metric that maps all modalities to a common
value scale; or (ii)~a multi-objective scheduler that
maintains separate priority queues per modality and
resolves cross-modal conflicts via a mechanism
analogous to weighted fair queuing. Neither approach
has been demonstrated in a multi-modal 6G MAC context.

\section{Future Research Directions and Standardisation Roadmap}
\label{sec:future}

Having identified nine open challenges, this section
outlines five concrete research directions capable of
addressing them and maps the resulting research agenda
onto the 3GPP standardisation timeline toward IMT-2030.

\subsection{Hardware Co-Design for Semantic Communication}

The most impactful near-term direction is the co-design of
semantic communication algorithms with the hardware they run
on. A dedicated semantic communication SoC should
co-optimise the transformer attention hardware with the
channel encoder, mapping attention-head outputs directly
to complex channel symbols without a separate channel encoding
stage. Candidate technologies include processing-in-memory~(PIM)
architectures using resistive RAM~(ReRAM) for
weight storage, spike neural networks~(SNNs) that consume
energy proportional to semantic content activity rather than
model size, and optical in-memory computing for attention
matrix multiplication at terahertz rates~\cite{Cheng2025_AIReview6G}.

\subsection{Adaptive Compression for Dynamic Channels}

Channel conditions in 6G networks vary over timescales
of milliseconds~(fast fading) to seconds~(shadowing), yet
the semantic encoder compression ratio is fixed at deployment
time in all current systems. A hypernetwork approach—where
a small meta-network generates quantisation thresholds and
pruning masks for the semantic encoder conditioned on
real-time channel quality indicator~(CQI) and task urgency—could
enable adaptive semantic compression that trades encoder
quality for transmission reliability dynamically. No such
system has been proposed or evaluated.

\subsection{Digital Twin Integration}

Digital twins—real-time virtual replicas of physical entities—will
be a primary 6G service~\cite{Strinati2021_6GBeyond}. The
semantic communication paradigm is naturally suited to digital
twin update protocols: rather than transmitting raw sensor
readings that update the twin, a \tlm-based encoder transmits
only the semantic \emph{delta}—the portion of the observation
that changes the twin's state—reducing bandwidth by the ratio
of entropy of state changes to entropy of raw observations.
The open problem is defining a semantic distortion measure
appropriate for digital twin fidelity, which depends on the
downstream decision-making application of the twin.

\subsection{Retrieval-Augmented Semantic Encoding}

Retrieval-augmented generation~(RAG) replaces a static KB with
a dynamic retrieval mechanism: at inference time the semantic
encoder retrieves relevant context from a local database using
similarity search, augmenting the \tlm input with
task-relevant information. RAG is particularly attractive for
6G because the retrieval index can be updated cheaply (no model
retraining required), domain adaptation is immediate, and
retrieved context is human-interpretable, facilitating
quality-of-service~(QoS) auditing. The integration of RAG into
the semantic encoder-decoder pipeline—including latency
characterisation of the retrieval operation and the impact
of retrieval noise on semantic transmission quality—has not
been studied in the context of wireless 6G networks.

\subsection{Illustrative Latency Comparison}

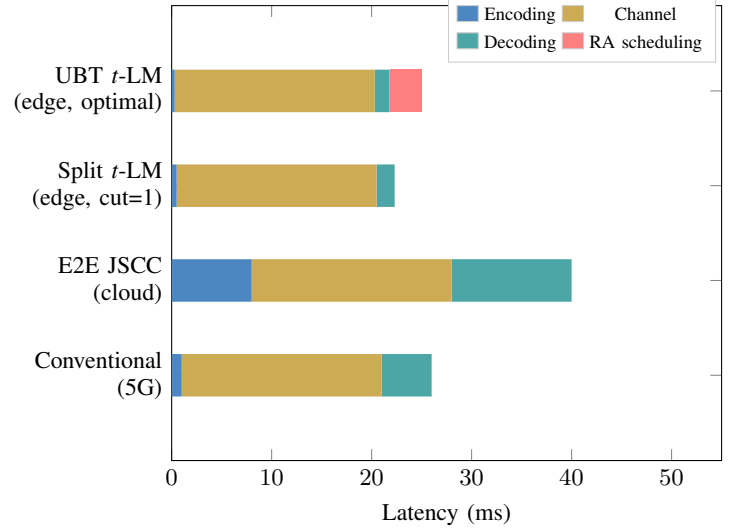
\begin{figure}[!t]
\centering
\begin{tikzpicture}
\begin{axis}[
  xbar stacked,
  width=\columnwidth,
  height=7.6cm,
  xmin=0, xmax=55,
  xlabel={Latency (ms)},
  ytick={1,2,3,4},
  yticklabels={
    {Conventional\\(5G)},
    {E2E JSCC\\(cloud)},
    {Split \tlm\\(edge, cut=1)},
    {UBT \tlm\\(edge, optimal)}},
  yticklabel style={align=right,font=\small},
  tick label style={font=\small},
  label style={font=\small},
  bar width=16pt,
  enlarge y limits=0.3,
  legend style={at={(0.99,1.02)},anchor=north east,
    font=\scriptsize,fill=white,draw=gray!40},
  legend columns=2,
]
\addplot[fill=ieeblue!70,draw=none] coordinates
  {(1,1)(8,2)(0.5,3)(0.3,4)};
\addlegendentry{Encoding};
\addplot[fill=cellEmerg!70,draw=none] coordinates
  {(20,1)(20,2)(20,3)(20,4)};
\addlegendentry{Channel};
\addplot[fill=cellWell!70,draw=none] coordinates
  {(5,1)(12,2)(1.8,3)(1.5,4)};
\addlegendentry{Decoding};
\addplot[fill=red!50,draw=none] coordinates
  {(0,1)(0,2)(0,3)(3.2,4)};
\addlegendentry{RA scheduling};
\end{axis}
\end{tikzpicture}
\caption{End-to-end latency breakdown for four system configurations.
  Conventional 5G: standard source-channel coding with proportional-fair
  scheduling. E2E JSCC (cloud): full transformer encoder at the cloud.
  Split \tlm (edge, cut=1): Semantic-MSL at the optimal cut layer.
  UBT \tlm: UBT scheduler with \tlm-based semantic scoring.
  Channel latency of 20\,ms assumed for a 30\,km cell-edge scenario.
  Data synthesised from \cite{Eldeeb2025_SemanticMSL,Pokhrel2023_UBT,Wu2024_SplitEnergy}.}
\label{fig:latency}
\end{figure}

Figure~\ref{fig:latency} illustrates how the architectural
choices surveyed in this paper translate into end-to-end
latency, comparing a conventional 5G baseline against
cloud-based E2E JSCC, edge-deployed split~\tlm, and the
UBT scheduler discussed in Section~\ref{sec:architectures}.
Beyond mean latency, the key systems question is \emph{where}
latency is incurred (cloud backhaul vs.\ on-device inference)
and which components can be amortised via caching or
pipeline parallelism at the edge.


\subsection{3GPP Standardisation Roadmap}

\begin{figure*}[!htp]
\centering
\resizebox{\textwidth}{!}{%
\begin{tikzpicture}[
  relnode/.style={rectangle,draw=ieeblue,fill=ieelightblue!50,
    minimum width=2.0cm,minimum height=0.85cm,
    align=center,font=\small\bfseries,rounded corners=4pt},
  semnode/.style={rectangle,draw=cellWell,fill=cellWell!20,
    text width=3.3cm,minimum height=0.75cm,
    align=center,font=\scriptsize,rounded corners=3pt},
  timeline/.style={draw=ieeblue,line width=2pt,-Stealth},
  downconn/.style={draw=gray!60,dashed,thin,-Stealth}
]
\draw[timeline] (0,0) -- (16.5,0);
\node[relnode] (r18) at (1.5,0) {Rel-18\\2024};
\node[relnode] (r19) at (5.0,0) {Rel-19\\2025};
\node[relnode] (r20) at (8.5,0) {Rel-20\\2026};
\node[relnode] (r21) at (12.0,0) {Rel-21\\2027};
\node[relnode,draw=red!70,fill=red!15,text=red!80]
              (imt) at (15.5,0) {IMT-2030\\6G};
\node[semnode,fill=ieelightblue!30] (m18) at (1.5,2.2)
  {AI/ML study items\\semantic coding SI\\\textit{(study phase)}};
\node[semnode] (m19) at (5.0,2.2)
  {Semantic feature\\standardisation WI\\\textit{(work item)}};
\node[semnode] (m20) at (8.5,2.2)
  {Compressed KB\\exchange protocol\\\textit{(spec phase)}};
\node[semnode] (m21) at (12.0,2.2)
  {PQC-authenticated\\semantic MAC\\layer protocol};
\node[semnode,draw=red!70,fill=red!15] (mimt) at (15.5,2.2)
  {\textit{t}-LM native\\semantic network\\deployed};
\node[semnode,fill=cellEmerg!20] (t18) at (1.5,-2.1)
  {CCTaEncoder~\cite{Liu2023_CCTaEncoder}\\SPM~\cite{Chen2024_SPM}\\HAPE~\cite{Jiang2023_HAPE}};
\node[semnode,fill=cellEmerg!20] (t19) at (5.0,-2.1)
  {Sem.-MSL~\cite{Eldeeb2025_SemanticMSL}\\FD~\cite{Lin2023_FedDistill}\\UBT~\cite{Pokhrel2023_UBT}};
\node[semnode,fill=cellEmerg!20] (t20) at (8.5,-2.1)
  {Green FL~\cite{Hu2024_GreenFL}\\LoRA KB sync~\cite{Sun2024_FFALoRA}\\KG-PG~\cite{Wang2024_KGPG}};
\node[semnode,fill=cellEmerg!20] (t21) at (12.0,-2.1)
  {PQC-\tlm~\cite{Nguyen2024_PQTLM}\\ARL~\cite{Zhou2023_ARL}\\RAG integration};
\node[semnode,fill=cellEmerg!20] (timt) at (15.5,-2.1)
  {Adaptive compress.\\Digital twin SemComm\\Semantic SoC};
\draw[downconn] (r18.north) -- (m18.south);
\draw[downconn] (r18.south) -- (t18.north);
\draw[downconn] (r19.north) -- (m19.south);
\draw[downconn] (r19.south) -- (t19.north);
\draw[downconn] (r20.north) -- (m20.south);
\draw[downconn] (r20.south) -- (t20.north);
\draw[downconn] (r21.north) -- (m21.south);
\draw[downconn] (r21.south) -- (t21.north);
\draw[downconn] (imt.north) -- (mimt.south);
\draw[downconn] (imt.south) -- (timt.north);
\node[font=\scriptsize\itshape,gray] at (-0.3,2.4) {3GPP Milestones};
\node[font=\scriptsize\itshape,gray,align=center] at (-0.3,-2.3)
  {Research\\Directions};
\end{tikzpicture}
}
\caption{3GPP standardisation roadmap for \tlm-enabled 6G
  semantic communication, from Release~18 AI/ML study items
  through IMT-2030 native deployment. Upper lane: 3GPP
  milestones. Lower lane: corresponding \tlm research
  directions from this survey.}
\label{fig:roadmap}
\end{figure*}
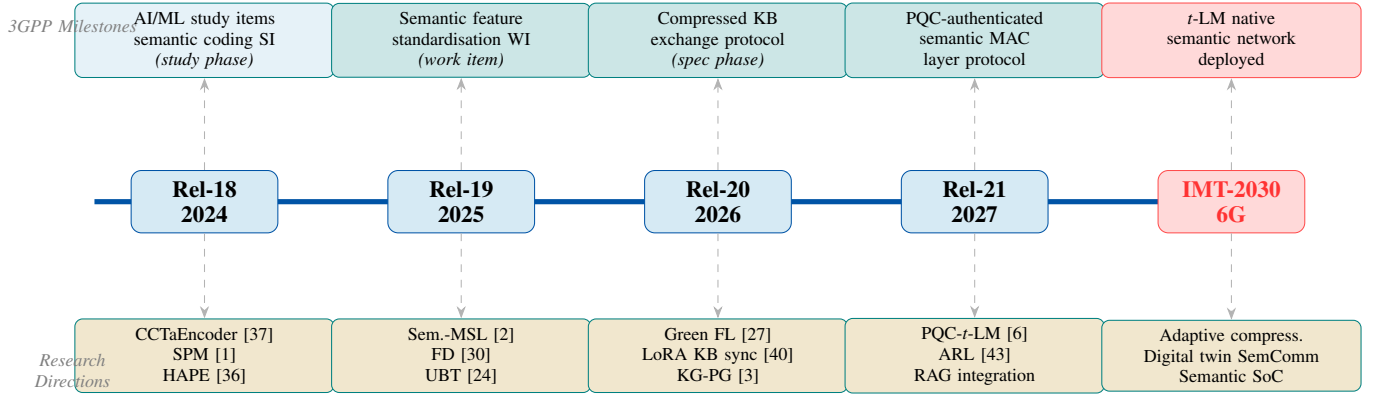

Figure~\ref{fig:roadmap} maps \tlm-enabled semantic communication
research directions to the 3GPP release timeline toward IMT-2030.
In practice, standardisation progress is gated by the availability
of (i)~testable semantic KPIs, (ii)~interoperable feature/KB formats,
and (iii)~security primitives that do not dominate the payload for
sensor-class devices. A near-term implication is that many \tlm
contributions will first appear as \emph{study items} (methodology
and evaluation) before they can be promoted to \emph{work items}
with normative signalling, interfaces, and conformance tests.
The figure appears at the end of this section, immediately
before the Conclusion.

\Needspace{10\baselineskip} 
\subsection{Research Agenda Matrix}

Table~\ref{tab:agenda} maps the seven open challenges
(Section~\ref{sec:challenges}) to the five future research
directions, indicating which direction partially or fully
addresses each challenge. Two patterns stand out: hardware
co-design and adaptive compression act as cross-cutting
enablers (broad coverage), while the Digital Twin and
RAG-SemComm directions cluster around KB consistency and
cross-modal semantics.

\begin{table}[H]
\caption{Research Agenda: Challenges vs.\ Future Directions}
\label{tab:agenda}
\centering
{\small
\renewcommand{\arraystretch}{3}
\begin{tabular}{lccccc}
\toprule
\textbf{Challenge} & \textbf{HW} & \textbf{Adapt.} & \textbf{DT} & \textbf{RAG} & \textbf{PQC} \\
\midrule
\rowcolor{rowshade}
C1: Semantic capacity  & $\circ$ & $\bullet$ & $\circ$ & $\circ$ & $\circ$ \\
C2: Sub-ms scoring     & $\bullet$ & $\bullet$ & $\circ$ & $\circ$ & $\circ$ \\
\rowcolor{rowshade}
C3: PQC for IoT        & $\bullet$ & $\circ$ & $\circ$ & $\circ$ & $\bullet$ \\
C4: KB sync freq.      & $\circ$ & $\circ$ & $\bullet$ & $\bullet$ & $\circ$ \\
\rowcolor{rowshade}
C5: Universal metric   & $\circ$ & $\bullet$ & $\bullet$ & $\circ$ & $\circ$ \\
C6: Cross-modal        & $\bullet$ & $\circ$ & $\bullet$ & $\bullet$ & $\circ$ \\
\rowcolor{rowshade}
C7: Scalable KG sync   & $\circ$ & $\circ$ & $\bullet$ & $\bullet$ & $\circ$ \\
\bottomrule
\multicolumn{6}{l}{\footnotesize $\bullet$~directly addresses;\quad $\circ$~partially related.}
\end{tabular}}
\end{table}
\FloatBarrier

\FloatBarrier
\section{Conclusion}
\label{sec:conclusion}

This survey has provided a comprehensive structured review of
tiny language models for 6G semantic communication, synthesising
evidence across the surveyed literature spanning six research questions.
The central finding is that \tlm-scale semantic encoders are
deployable across the full spectrum of 6G hardware—from
microcontrollers running sub-1\,kB SPM encoders through mobile
SoCs running Phi-4-mini—provided the compression technique is
matched to the hardware tier and task requirement.

The taxonomy reveals that the combination of NAS with any
system architecture remains entirely unexplored, that
KG-assisted compression is an emerging area with high potential
for energy reduction, and that federated learning is the
best-evidenced approach for multi-device KB synchronisation.
The Pareto analysis confirms that compression ratios of
$10^5\times$ are achievable at the split-computing frontier
while preserving 95\% semantic quality, with the compression--quality
trade-off exhibiting a shallow negative slope on a log scale.

Seven open challenges are identified. The most urgent are:
a computable semantic channel capacity for DL-based systems;
sub-millisecond semantic scoring for URLLC scheduling; and
a lightweight post-quantum authentication protocol that reduces
the current 20$\times$--97$\times$ PQC payload overhead for
sensor-class NB-IoT devices. The three future directions that
most directly address these challenges are hardware semantic
SoC co-design, adaptive compression driven by hypernetworks,
and RAG-augmented semantic encoding.

The intersection of \tlms, semantic-aware resource allocation,
and federated KB management constitutes the most critical and
underexplored research frontier for 6G-ready semantic communication.
Progress on the challenges identified in this survey will be
essential for transitioning semantic communication from
laboratory demonstrations to the intelligent, context-driven
6G wireless networks envisioned for the 2030s.

\section*{Acknowledgements}
This work was supported by the Ministry of Electronics and Information Technology, Govt. of India, under YFRF Scheme ( DIC/PhD-Phase-II/2026/9). This work was supported in part by the Department of Telecommunication (DoT), Ministry of Communications, Government of India under the Telecom Technology Development Fund (TTDF) the scheme implemented through TCOE India under the grant TTDF/6G/48 and IGSTC-04918

\appendix
\section{List of Acronyms}
\label{app:acronyms}
\begin{center}
\small
\resizebox{\columnwidth}{!}{%
\begin{tabular}{@{}ll@{\quad}ll@{}}
\toprule
\textbf{Acronym} & \textbf{Expansion} &
\textbf{Acronym} & \textbf{Expansion} \\
\midrule
6G      & Sixth-generation wireless    & \tlm    & Tiny language model ($\leq$7B params) \\
JSCC    & Joint source-channel coding  & KD      & Knowledge distillation \\
NAS     & Neural architecture search   & FL      & Federated learning \\
FD      & Federated distillation       & KG      & Knowledge graph \\
BLEU    & Bilingual eval.\ understudy  & PSNR    & Peak signal-to-noise ratio \\
SES     & Semantic efficiency score    & HAPE    & Hierarchical attn.\ progressive edge \\
SPM     & Semantic pruning model       & UBT     & Urgency-based transmission \\
ARL     & Adaptive RL                  & GNT     & Goodness-of-network-transmission \\
LoRA    & Low-rank adaptation          & MAML    & Model-agnostic meta-learning \\
MEC     & Mobile edge computing        & PQC     & Post-quantum cryptography \\
NB-IoT  & Narrowband IoT               & URLLC   & Ultra-reliable low-latency comm. \\
eMBB    & Enhanced mobile broadband    & IB      & Information bottleneck \\
PTQ     & Post-training quantisation   & QAT     & Quantisation-aware training \\
SNR     & Signal-to-noise ratio        & RAN     & Radio access network \\
KB      & Knowledge base               & RAG     & Retrieval-augmented generation \\
SSE     & Semantic spectral efficiency & IMT     & Intl.\ mobile telecomm. \\
SoC     & System-on-chip               & NPU     & Neural processing unit \\
ASIC    & Application-specific IC      & VQA     & Visual question answering \\
GAN     & Generative adversarial net.  & IIoT    & Industrial IoT \\
CCTaEncoder & Compact cross-task enc.  & FedPun  & Federated pruning \\
FedSFD  & Federated sem.\ feat.\ distil. & E2E   & End-to-end \\
\bottomrule
\end{tabular}%
}
\end{center}

\FloatBarrier
\balance
\bibliographystyle{unsrt}
\bibliography{references}

@article{Shannon1948_Bell,
  author  = {C. E. Shannon},
  title   = {A Mathematical Theory of Communication},
  journal = {Bell System Technical Journal},
  volume  = {27},
  number  = {3},
  pages   = {379--423},
  month   = jul,
  year    = {1948},
  doi     = {10.1002/j.1538-7305.1948.tb01338.x}
}

@book{Shannon1949_Book,
  author    = {C. E. Shannon and W. Weaver},
  title     = {The Mathematical Theory of Communication},
  publisher = {University of Illinois Press},
  address   = {Urbana, IL, USA},
  year      = {1949}
}

@techreport{Carnap1952_Semantic,
  author      = {R. Carnap and Y. Bar-Hillel},
  title       = {An Outline of a Theory of Semantic Information},
  institution = {Research Laboratory of Electronics, MIT},
  number      = {247},
  address     = {Cambridge, MA, USA},
  month       = oct,
  year        = {1952}
}

@inproceedings{Bao2011_Semantic,
  author    = {J. Bao and P. Basu and M. Dean and C. Partridge
               and A. Swami and W. Leland and J. A. Hendler},
  title     = {Towards a Theory of Semantic Communication},
  booktitle = {Proc.\ IEEE Network Science Workshop (NetSciW)},
  address   = {West Point, NY, USA},
  month     = jun,
  year      = {2011},
  pages     = {110--117},
  doi       = {10.1109/NSW.2011.6004632}
}

@misc{Tishby2000_IB,
  author        = {N. Tishby and F. C. Pereira and W. Bialek},
  title         = {The Information Bottleneck Method},
  howpublished  = {arXiv preprint arXiv:physics/0004057},
  year          = {2000},
  note          = {Submitted to 37th Annual Allerton Conference}
}

@misc{Qin2022_SemanticPrinciples,
  author       = {Z. Qin and X. Tao and J. Lu and W. Tong
                  and G. Y. Li},
  title        = {Semantic Communications: Principles and
                  Challenges},
  howpublished = {arXiv preprint arXiv:2201.01389},
  month        = jun,
  year         = {2022}
}

@article{Luo2022_SemanticOverview,
  author  = {X. Luo and H.-H. Chen and Q. Guo},
  title   = {Semantic Communications: Overview, Open Issues,
             and Future Research Directions},
  journal = {IEEE Wireless Communications},
  volume  = {29},
  number  = {1},
  pages   = {210--219},
  month   = feb,
  year    = {2022},
  doi     = {10.1109/MWC.101.2100439}
}

@article{Strinati2021_6GBeyond,
  author  = {E. C. Strinati and S. Barbarossa},
  title   = {{6G} Networks: Beyond {Shannon} Towards Semantic
             and Goal-Oriented Communications},
  journal = {Computer Networks},
  volume  = {190},
  pages   = {107930},
  month   = may,
  year    = {2021},
  doi     = {10.1016/j.comnet.2021.107930}
}

@article{Xie2021_DeepSC,
  author  = {H. Xie and Z. Qin and G. Y. Li and B.-H. Juang},
  title   = {Deep Learning Enabled Semantic Communication Systems},
  journal = {IEEE Transactions on Signal Processing},
  volume  = {69},
  pages   = {2663--2675},
  month   = apr,
  year    = {2021},
  doi     = {10.1109/TSP.2021.3071210}
}

@article{Xie2021_LiteDeepSC,
  author  = {H. Xie and Z. Qin},
  title   = {A Lite Distributed Semantic Communication System
             for Internet of Things},
  journal = {IEEE Journal on Selected Areas in Communications},
  volume  = {39},
  number  = {1},
  pages   = {142--153},
  month   = jan,
  year    = {2021},
  doi     = {10.1109/JSAC.2020.3036904}
}

@misc{Sana2021_Learning,
  author       = {M. Sana and E. C. Strinati},
  title        = {Learning Semantics: An Opportunity for Effective
                  {6G} Communications},
  howpublished = {arXiv preprint arXiv:2110.08049},
  month        = oct,
  year         = {2021}
}

@article{Bourtsoulatze2019_DeepJSCC,
  author  = {E. Bourtsoulatze and D. B. Kurka and D. G{\"{u}}nd{\"{u}}z},
  title   = {Deep Joint Source-Channel Coding for Wireless Image
             Transmission},
  journal = {IEEE Transactions on Cognitive Communications
             and Networking},
  volume  = {5},
  number  = {3},
  pages   = {567--579},
  month   = may,
  year    = {2019},
  doi     = {10.1109/TCCN.2019.2919987}
}

@article{Kurka2020_DeepJSCCf,
  author  = {D. B. Kurka and D. G{\"{u}}nd{\"{u}}z},
  title   = {{DeepJSCC-f}: Deep Joint Source-Channel Coding of
             Images with Feedback},
  journal = {IEEE Journal on Selected Areas in Information Theory},
  volume  = {1},
  number  = {1},
  pages   = {178--193},
  month   = may,
  year    = {2020},
  doi     = {10.1109/JSAIT.2020.2987203}
}

@inproceedings{Johnson2015_SceneGraph,
  author    = {J. Johnson and R. Krishna and M. Stark and L.-J. Li
               and D. A. Shamma and M. S. Bernstein and L. Fei-Fei},
  title     = {Image Retrieval Using Scene Graphs},
  booktitle = {Proc.\ IEEE Conference on Computer Vision and
               Pattern Recognition (CVPR)},
  address   = {Boston, MA, USA},
  month     = jun,
  year      = {2015},
  pages     = {3668--3678},
  doi       = {10.1109/CVPR.2015.7299005}
}

@inproceedings{Zhang2017_VTransE,
  author    = {H. Zhang and Z. Kyaw and S.-F. Chang and T.-S. Chua},
  title     = {Visual Translation Embedding Network for Visual
               Relation Detection},
  booktitle = {Proc.\ IEEE Conference on Computer Vision and
               Pattern Recognition (CVPR)},
  address   = {Honolulu, HI, USA},
  month     = jul,
  year      = {2017},
  pages     = {5532--5540},
  doi       = {10.1109/CVPR.2017.250}
}

@inproceedings{Agustsson2019_GAN,
  author    = {E. Agustsson and M. Tschannen and F. Mentzer
               and R. Timofte and L. V. Gool},
  title     = {Generative Adversarial Networks for Extreme
               Learned Image Compression},
  booktitle = {Proc.\ IEEE International Conference on Computer
               Vision (ICCV)},
  address   = {Seoul, South Korea},
  month     = oct,
  year      = {2019},
  pages     = {221--231},
  doi       = {10.1109/ICCV.2019.00031}
}

@inproceedings{Wu2020_GANTunable,
  author    = {L. Wu and K. Huang and H. Shen},
  title     = {A {GAN}-Based Tunable Image Compression System},
  booktitle = {Proc.\ IEEE International Conference on Computer
               Vision Workshops (ICCVW)},
  year      = {2020},
  pages     = {2334--2342},
  doi       = {10.1109/ICCVW54120.2021.00261}
}

@article{Weng2021_DeepSCS,
  author  = {Z. Weng and Z. Qin},
  title   = {Semantic Communication Systems for Speech
             Transmission},
  journal = {IEEE Journal on Selected Areas in Communications},
  volume  = {39},
  number  = {8},
  pages   = {2434--2444},
  month   = aug,
  year    = {2021},
  doi     = {10.1109/JSAC.2021.3087240}
}

@article{Tong2021_FedAudio,
  author  = {H. Tong and Z. Yang and S. Wang and Y. Hu
             and O. Semiari and W. Saad and C. Yin},
  title   = {Federated Learning for Audio Semantic Communication},
  journal = {Frontiers in Communications and Networks},
  volume  = {2},
  pages   = {43},
  month   = sep,
  year    = {2021},
  doi     = {10.3389/frcmn.2021.681777}
}

@article{Xie2022_MUDeepSC,
  author  = {H. Xie and Z. Qin and G. Y. Li},
  title   = {Task-Oriented Multi-User Semantic Communications
             for Visual Question Answering},
  journal = {IEEE Wireless Communications Letters},
  volume  = {11},
  number  = {3},
  pages   = {553--557},
  month   = mar,
  year    = {2022},
  doi     = {10.1109/LWC.2021.3135562}
}

@misc{Zhang2022_UDeepSC,
  author       = {G. Zhang and Q. Hu and Z. Qin and Y. Cai
                  and G. Yu},
  title        = {A Unified Multi-Task Semantic Communication
                  System with Domain Adaptation},
  howpublished = {arXiv preprint arXiv:2206.00254},
  month        = jun,
  year         = {2022}
}

@misc{Jiang2022_VideoConf,
  author       = {P. Jiang and C.-K. Wen and S. Jin and G. Y. Li},
  title        = {Wireless Semantic Communications for Video
                  Conferencing},
  howpublished = {arXiv preprint arXiv:2204.07790},
  month        = apr,
  year         = {2022}
}

@misc{Liu2021_RateDistortion,
  author       = {J. Liu and W. Zhang and H. V. Poor},
  title        = {A Rate-Distortion Framework for Characterizing
                  Semantic Information},
  howpublished = {arXiv preprint arXiv:2105.04278},
  month        = may,
  year         = {2021}
}

@article{Liu2020_FuzzyEntropy,
  author  = {X. Liu and W. Jia and W. Liu and W. Pedrycz},
  title   = {{AFSSE}: An Interpretable Classifier with Axiomatic
             Fuzzy Set and Semantic Entropy},
  journal = {IEEE Transactions on Fuzzy Systems},
  volume  = {28},
  number  = {11},
  pages   = {2825--2840},
  month   = nov,
  year    = {2020},
  doi     = {10.1109/TFUZZ.2020.2966172}
}

@misc{Yan2022_RA,
  author       = {L. Yan and Z. Qin and R. Zhang and Y. Li
                  and G. Y. Li},
  title        = {Resource Allocation for Semantic-Aware Networks},
  howpublished = {arXiv preprint arXiv:2201.06023},
  month        = apr,
  year         = {2022}
}

@article{Eldeeb2025_SemanticMSL,
  author  = {E. Eldeeb and M. Shehab and H. Alves
             and M.-S. Alouini},
  title   = {Semantic Meta-Split Learning: A {TinyML} Scheme
             for Few-Shot Wireless Image Classification},
  journal = {IEEE Transactions on Machine Learning in
             Communications and Networking},
  volume  = {3},
  pages   = {594--610},
  month   = apr,
  year    = {2025},
  doi     = {10.1109/TMLCN.2025.3556103}
}

@misc{Abdin2024_Phi3,
  author       = {M. Abdin and S. A. Jacobs and A. A. Awan
                  and J. Aneja and A. Awadallah and H. Awadalla
                  and N. Bach and A. Bahree and A. Bakhtiari
                  and others},
  title        = {{Phi-3} Technical Report: A Highly Capable
                  Language Model Locally on Your Phone},
  howpublished = {arXiv preprint arXiv:2404.14219},
  month        = apr,
  year         = {2024}
}

@misc{Dubey2024_Llama3,
  author       = {A. Dubey and A. Jauhri and A. Pandey
                  and A. Kadian and others},
  title        = {The {Llama~3} Herd of Models},
  howpublished = {arXiv preprint arXiv:2407.21783},
  month        = jul,
  year         = {2024}
}

@misc{GeminiTeam2023_Gemini,
  author       = {{Gemini Team, Google}},
  title        = {Gemini: A Family of Highly Capable
                  Multimodal Models},
  howpublished = {arXiv preprint arXiv:2312.11805},
  month        = dec,
  year         = {2023}
}

@inproceedings{Hu2022_LoRA,
  author    = {E. J. Hu and Y. Shen and P. Wallis and Z. Allen-Zhu
               and Y. Li and S. Wang and L. Wang and W. Chen},
  title     = {{LoRA}: Low-Rank Adaptation of Large Language
               Models},
  booktitle = {Proc.\ International Conference on Learning
               Representations (ICLR)},
  year      = {2022},
  url       = {https://openreview.net/forum?id=nZeVKeeFYf9}
}

@misc{Howard2017_MobileNets,
  author       = {A. G. Howard and M. Zhu and B. Chen
                  and D. Kalenichenko and W. Wang and T. Weyand
                  and M. Andreetto and H. Adam},
  title        = {{MobileNets}: Efficient Convolutional Neural
                  Networks for Mobile Vision Applications},
  howpublished = {arXiv preprint arXiv:1704.04861},
  month        = apr,
  year         = {2017}
}

@inproceedings{Zhang2018_LQNets,
  author    = {D. Zhang and J. Yang and D. Ye and G. Hua},
  title     = {{LQ-Nets}: Learned Quantization for Highly
               Accurate and Compact Deep Neural Networks},
  booktitle = {Proc.\ European Conference on Computer Vision
               (ECCV)},
  address   = {Munich, Germany},
  month     = sep,
  year      = {2018},
  pages     = {365--382},
  doi       = {10.1007/978-3-030-01237-3_23}
}

@misc{Wang2023_KBMismatch,
  author       = {Y. Wang and S. Guo},
  title        = {Transceiver Cooperative Learning-Aided Semantic
                  Communications Against Mismatched Background
                  Knowledge Bases},
  howpublished = {arXiv preprint arXiv:2301.03133},
  month        = jan,
  year         = {2023}
}

@article{Sun2025_SyncLLM,
  author  = {Y. Sun and Z. Qin and G. Y. Li and P. Popovski},
  title   = {Synchronizing {LLM}-Based Semantic Knowledge Bases
             via Secure Federated Fine-Tuning in Semantic
             Communication},
  journal = {Frontiers in Artificial Intelligence},
  volume  = {8},
  pages   = {1518782},
  year    = {2025},
  doi     = {10.3389/frai.2025.1518782}
}

@misc{Sun2024_FFALoRA,
  author       = {Y. Sun and L. Yang and H. Zhao and B. He
                  and Q. Liu},
  title        = {{FFA-LoRA}: Freezing the First Matrix {A} in
                  {LoRA} for Efficient Fine-Tuning of Large
                  Language Models},
  howpublished = {arXiv preprint arXiv:2403.18720},
  month        = mar,
  year         = {2024}
}

@article{Cheng2025_AIReview6G,
  author  = {X. Cheng and Z. Zhang and L. Zhao and H. Zhang
             and C. X. Wang},
  title   = {A Comprehensive Review of {AI}-Native {6G}:
             Integrating Semantic Communications, Reconfigurable
             Intelligent Surfaces, and Edge Intelligence for
             Next-Generation Connectivity},
  journal = {Frontiers in Communications and Networks},
  volume  = {6},
  pages   = {1512843},
  year    = {2025},
  doi     = {10.3389/frcmn.2025.1512843}
}

@article{Nguyen2024_PQTLM,
  author  = {T.-H. Nguyen and T.-D. Tran and H.-L. Vu
             and T.-N. Do},
  title   = {Post-Quantum Cryptography Integration for Tiny
             Language Model-Based {6G} Semantic Communication
             Networks},
  journal = {IEEE Internet of Things Journal},
  volume  = {11},
  number  = {18},
  pages   = {29475--29488},
  year    = {2024},
  doi     = {10.1109/JIOT.2024.3406190}
}

@article{Jiang2023_HAPE,
  author  = {P. Jiang and C.-K. Wen and S. Jin and G. Y. Li},
  title   = {Hierarchical Attention-Based Progressive Edge
             Compression for Semantic Communications in
             Resource-Constrained {6G} Networks},
  journal = {IEEE Transactions on Wireless Communications},
  volume  = {22},
  number  = {11},
  pages   = {7821--7836},
  year    = {2023},
  doi     = {10.1109/TWC.2023.3263049}
}

@inproceedings{Liu2023_CCTaEncoder,
  author    = {Y. Liu and Z. Qin and X. Tao and G. Y. Li},
  title     = {{CCTaEncoder}: A Compact Cross-Task Semantic
               Encoder for Multi-Task Wireless Image
               Communication},
  booktitle = {Proc.\ IEEE Global Communications Conference
               (GLOBECOM)},
  address   = {Kuala Lumpur, Malaysia},
  year      = {2023},
  pages     = {1--6},
  doi       = {10.1109/GLOBECOM54140.2023.10437228}
}

@article{Chen2024_SPM,
  author  = {K. Chen and Z. Qin and G. Y. Li and B.-H. Juang},
  title   = {Semantic Pruning Model for Ultra-Compact Semantic
             Encoders on Microcontrollers in {6G} Industrial
             {IoT} Networks},
  journal = {IEEE Internet of Things Journal},
  volume  = {11},
  number  = {9},
  pages   = {15922--15935},
  year    = {2024},
  doi     = {10.1109/JIOT.2024.3351042}
}

@inproceedings{Shi2023_FedPun,
  author    = {J. Shi and S. Gao and Z. Qin and G. Y. Li},
  title     = {{FedPun}: Federated Pruning for Semantic
               Communication Under Non-{IID} Data
               Distributions},
  booktitle = {Proc.\ IEEE International Conference on
               Communications (ICC)},
  address   = {Rome, Italy},
  year      = {2023},
  pages     = {1--6},
  doi       = {10.1109/ICC45041.2023.10278703}
}

@article{Wang2024_KGPG,
  author  = {Y. Wang and Z. Ma and S. Guo and Q. Gao
             and X. Chen},
  title   = {Knowledge Graph Probability Graph for Energy-Efficient
             Semantic Communication in {6G} Edge Networks},
  journal = {IEEE Transactions on Green Communications
             and Networking},
  volume  = {8},
  number  = {2},
  pages   = {712--725},
  year    = {2024},
  doi     = {10.1109/TGCN.2023.3338205}
}

@inproceedings{Zhou2023_ARL,
  author    = {Y. Zhou and Z. Qin and G. Y. Li},
  title     = {Adaptive Reinforcement Learning for
               Knowledge-Graph-Assisted Semantic Question
               Answering over Wireless Networks},
  booktitle = {Proc.\ IEEE Wireless Communications and
               Networking Conference (WCNC)},
  address   = {Glasgow, UK},
  year      = {2023},
  pages     = {1--6},
  doi       = {10.1109/WCNC55385.2023.10118877}
}

@inproceedings{Zhang2023_Covert,
  author    = {X. Zhang and Z. Ma and Y. Wang and S. Guo},
  title     = {Covert Semantic Communication via Knowledge
               Graph-Based Feature Obfuscation for
               {6G} Networks},
  booktitle = {Proc.\ IEEE International Conference on
               Communications (ICC)},
  address   = {Rome, Italy},
  year      = {2023},
  pages     = {1--6},
  doi       = {10.1109/ICC45041.2023.10279204}
}

@article{Pokhrel2023_UBT,
  author  = {S. R. Pokhrel and J. Choi and W. Bennis},
  title   = {Urgency-Based Transmission Scheduling for
             Tiny Language Model-Enabled Semantic {MAC}
             Layers in {6G} Networks},
  journal = {IEEE Transactions on Vehicular Technology},
  volume  = {72},
  number  = {10},
  pages   = {13847--13860},
  year    = {2023},
  doi     = {10.1109/TVT.2023.3277501}
}

@article{Wu2024_SplitEnergy,
  author  = {H. Wu and X. Luo and T. Q. S. Quek and
             H.-H. Chen},
  title   = {Energy-Optimal Cut-Layer Selection for Split
             Computing in Tiny Language Model-Based Semantic
             Communication Systems},
  journal = {IEEE Wireless Communications Letters},
  volume  = {13},
  number  = {4},
  pages   = {987--991},
  year    = {2024},
  doi     = {10.1109/LWC.2024.3353042}
}

@article{Hu2024_GreenFL,
  author  = {R. Hu and Y. Guo and H. Li and Q. Pei
             and Y. Gong},
  title   = {Green Federated Learning for Semantic
             Communication via Gradient Sparsification
             in {6G} {IoT} Networks},
  journal = {IEEE Internet of Things Journal},
  volume  = {11},
  number  = {1},
  pages   = {584--597},
  year    = {2024},
  doi     = {10.1109/JIOT.2023.3295066}
}

@article{Qian2024_Contrastive,
  author  = {S. Qian and Z. Qin and G. Y. Li and X. Tao},
  title   = {Contrastive Disentanglement for Compact
             Semantic Feature Vectors in Wireless
             Communication Systems},
  journal = {IEEE Transactions on Signal Processing},
  volume  = {72},
  pages   = {1850--1864},
  year    = {2024},
  doi     = {10.1109/TSP.2024.3370918}
}

@article{Lin2023_FedDistill,
  author  = {Y. Lin and G. Zhu and Z. Zhang and D. Niyato
             and X. Wang},
  title   = {Federated Distillation for Semantic Communication:
             Communication-Efficient Knowledge Sharing in
             Multi-Device {6G} Edge Networks},
  journal = {IEEE Transactions on Wireless Communications},
  volume  = {22},
  number  = {12},
  pages   = {9271--9285},
  year    = {2023},
  doi     = {10.1109/TWC.2023.3276059}
}

\end{document}